\documentclass[11pt]{article}
\usepackage{bbm}
\usepackage{hyperref}
\usepackage{longtable}
\usepackage{eqnarray,amsmath,amsfonts,amsthm,mathrsfs}
\usepackage{amssymb}
\usepackage{bm}
\usepackage{bbm}
\usepackage{amssymb}
\usepackage{mathtools}
\usepackage{mathrsfs}
\usepackage{physics}
\usepackage{bbm}
\usepackage{algorithm}

\usepackage{algorithmic}
\usepackage{color}
\usepackage{bm}
\usepackage{amssymb}
\usepackage{graphicx}
\usepackage{epsfig}
\usepackage{epsf}
\usepackage{float}
\usepackage{subfigure}
\usepackage{amsfonts,amsmath,amsthm,amssymb,graphicx,float,fancyhdr,multirow,hyperref}
\usepackage{booktabs,longtable,authblk}
\usepackage{mathrsfs,hhline}
\usepackage{makecell}
\usepackage{fancyhdr}
\usepackage[top=2.5cm, bottom=2.5cm, left=3cm, right=3cm]{geometry}
\usepackage{graphicx}
\newtheorem{theorem}{Theorem}

\newtheorem{lemma}{Lemma}

\newtheorem{proof of lemma}{Proof of Lemma}[section]

\allowdisplaybreaks[4]

\numberwithin{equation}{section}

\title{Mining and Intervention of Social Networks Information Cocoon Based on Multi-Layer Network Community Detection$^\dag$\footnotetext{\dag~The corresponding author is Lei Shi.}}

\author[1]{Suwen Yang}
\author[1,2]{Lei Shi}

\affil[1]{School of Mathematical Sciences, \linebreak
	Fudan University, Shanghai, 200433, China }
\affil[2]{ Shanghai Key Laboratory for Contemporary Applied Mathematics, \linebreak
	Fudan University, Shanghai, 200433, China \linebreak
	Email:swyang21@m.fudan.edu.cn, leishi@fudan.edu.cn}

\date{}

\begin{document}
	\maketitle
\begin{abstract}
With the rapid development of information technology and the widespread utilization of recommendation algorithms, users are able to access information more conveniently, while the content they receive tends to be homogeneous. Homogeneous viewpoints and preferences tend to cluster users into sub-networks, leading to group polarization and increasing the likelihood of forming information cocoons. This paper aims to handle information cocoon phenomena in debates on social media. In order to investigate potential user connections, we construct a double-layer network that incorporates two dimensions: relational ties and feature-based similarity between users. Based on the structure of the multi-layer network, we promote two graph auto-encoder (GAE) based community detection algorithms, which can be applied to the partition and determination of information cocoons. This paper tests these two algorithms on Cora, Citeseer, and synthetic datasets, comparing them with existing multi-layer network unsupervised community detection algorithms. Numerical experiments illustrate that the algorithms proposed in this paper significantly improve prediction accuracy indicator NMI (normalized mutual information) and network topology indicator Q. Additionally, an influence-based intervention measure on which algorithms can operate is proposed. Through the Markov states transition model, we simulate the intervention effects, which illustrate that our community detection algorithms play a vital role in partitioning and determining information cocoons. Simultaneously, our intervention strategy alleviates the polarization of viewpoints and the formation of information cocoons with minimal intervention effort.
		
		
\end{abstract}
	
{\textbf{Keywords and phrases:} Multi-layer social network; Community detection; Information cocoon; Modularity tensor reconstruction; Graph auto-encoder (GAE); Markov state transition model.}

\section{Introduction}\label{Section: Introduction}
With the exponential growth of the Internet and the ubiquitous utilization of recommendation algorithms, the information cocoon phenomenon has gained increasing traction recently. As information companies continuously analyze users’ preferences and refine recommendation algorithms to maximize business profits, individuals also benefit from increasingly personalized information services.
While rapid development of information technology improves the experience of users, it inevitably leads to users exposing to homogeneous information, which may contribute to information narrowing and group polarization, ultimately resulting in the formation of information cocoons \cite{sunstein2006infotopia}. 

The concept of information cocoons was first proposed in ``Infotopia: How Many Minds Produce Knowledge'' by Sunstein in 2006 \cite{sunstein2006infotopia}. People's attention would instinctively focus on those topics where their interests lay and tend to search for ideas that could support their standpoints. As for those viewpoints that are opposite from their standpoints, they could choose to ignore them automatically. Information cocoons are warm and friendly spaces for people living in an information space where people only accept the views they like and exclude information they oppose \cite{sunstein2006infotopia}. Although it seems natural for individuals to search the viewpoints where their interests lie, there has been extensive research demonstrating that personalized recommendations create filter bubbles, thus forcing the occurrence of information cocoon phenomena \cite{piao2023human}, \cite{hou2023information}, \cite{pariser2011filter}.

\begin{figure}[H]
	\centering
	\begin{minipage}[b]{0.75\textwidth}
		\includegraphics[width=\textwidth,height=0.45\textwidth]{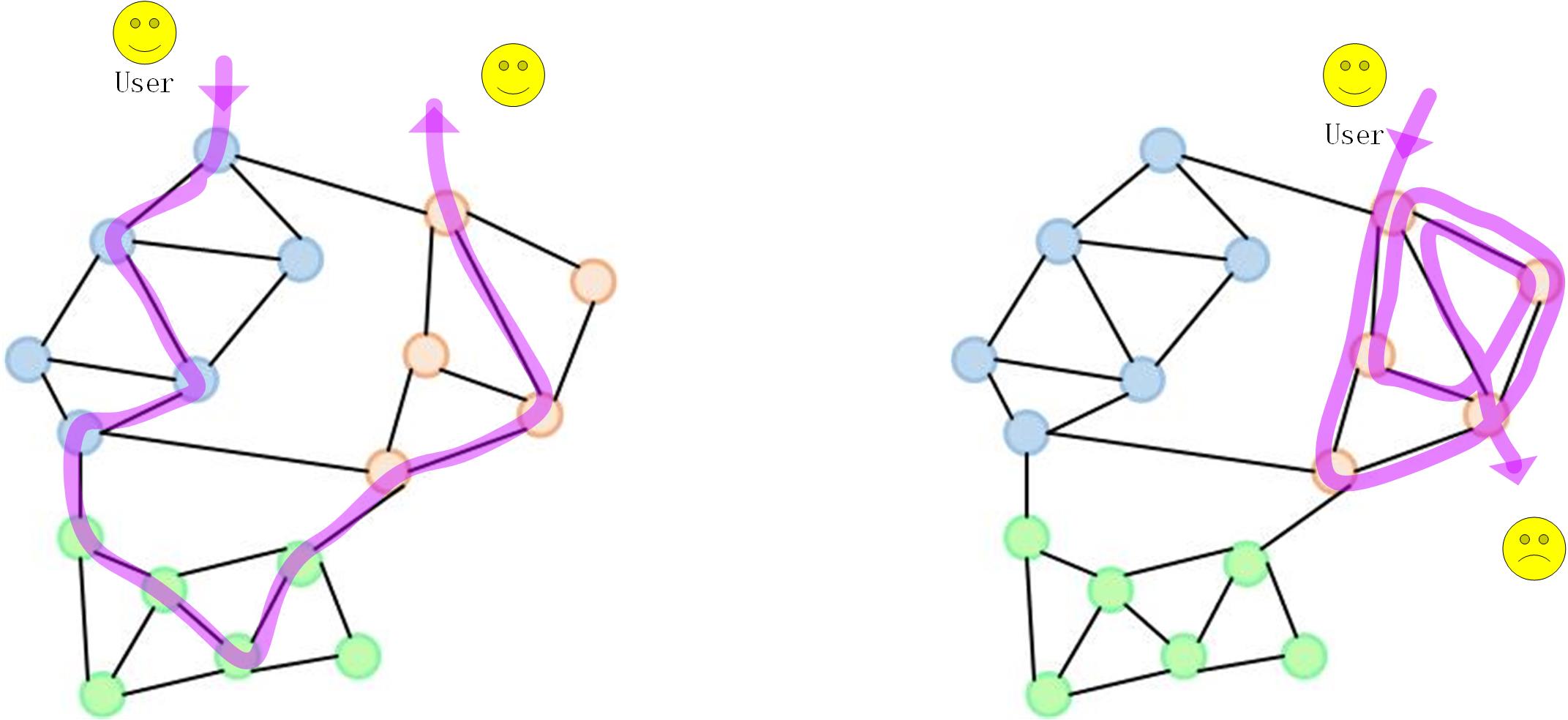}
		\caption{Browsing paths with or without a recommendation system}
		\label{fig:information cocoons formation under recommendation system}
	\end{minipage}
\end{figure}

Without recommendation systems, the search process is time-consuming and costly, requiring individuals to explore the entire event framework on their own. Under such circumstances, people can obtain a comprehensive understanding of events, thus being able to view issues objectively. The left of \autoref{fig:information cocoons formation under recommendation system} depicts the searching tracks of users under non-recommendation conditions. Under the recommendation system, internet technology operators could precisely locate where certain users are interested, allowing users to browse what they want to obtain directly without seeing much redundant information. 
The right part of \autoref{fig:information cocoons formation under recommendation system} depicts the approximate browsing trajectory of users under the influence of recommendation algorithms.

In general, the widely recognized contributing factors to the formation of information cocoons mainly stem from two primary sources: information delivery mechanisms and personal information preferences \cite{gu2024modeling}. Obviously, under the influence of data-driven algorithms, the impact of the former appears to be more severe. Recommendation algorithms not only make people lack comprehensive command of the whole topics, contributing to extreme views more easily generated but also reduce the browsing time for people to find their interested points, leading to information cocoons accelerated forming \cite{piao2023human}. It has been suggested that homogeneous and singular information aggravate social opinion polarization and breed extremism. When people find their opinion supported by others, they will be confident to become more extreme, thus accelerating the dissemination of radical emotions and even group conflicts \cite{sunstein2019trusting}. Take the elections in 2012 and 2016 as examples: social media plays a guiding role in shaping public opinions, which not only contributed to an inducing impact on the final election results to some degree but also triggered offline conflicts in multiple states \cite{sunstein2018republic}, \cite{guess2023social}. Therefore, it gradually becomes a high-profile issue to excavate the information cocoons in time and take intervention measures making full use of the dissemination of internet information.

\begin{figure}[H]
	\centering
	\begin{minipage}[b]{0.6\textwidth}
		\includegraphics[width=\textwidth]{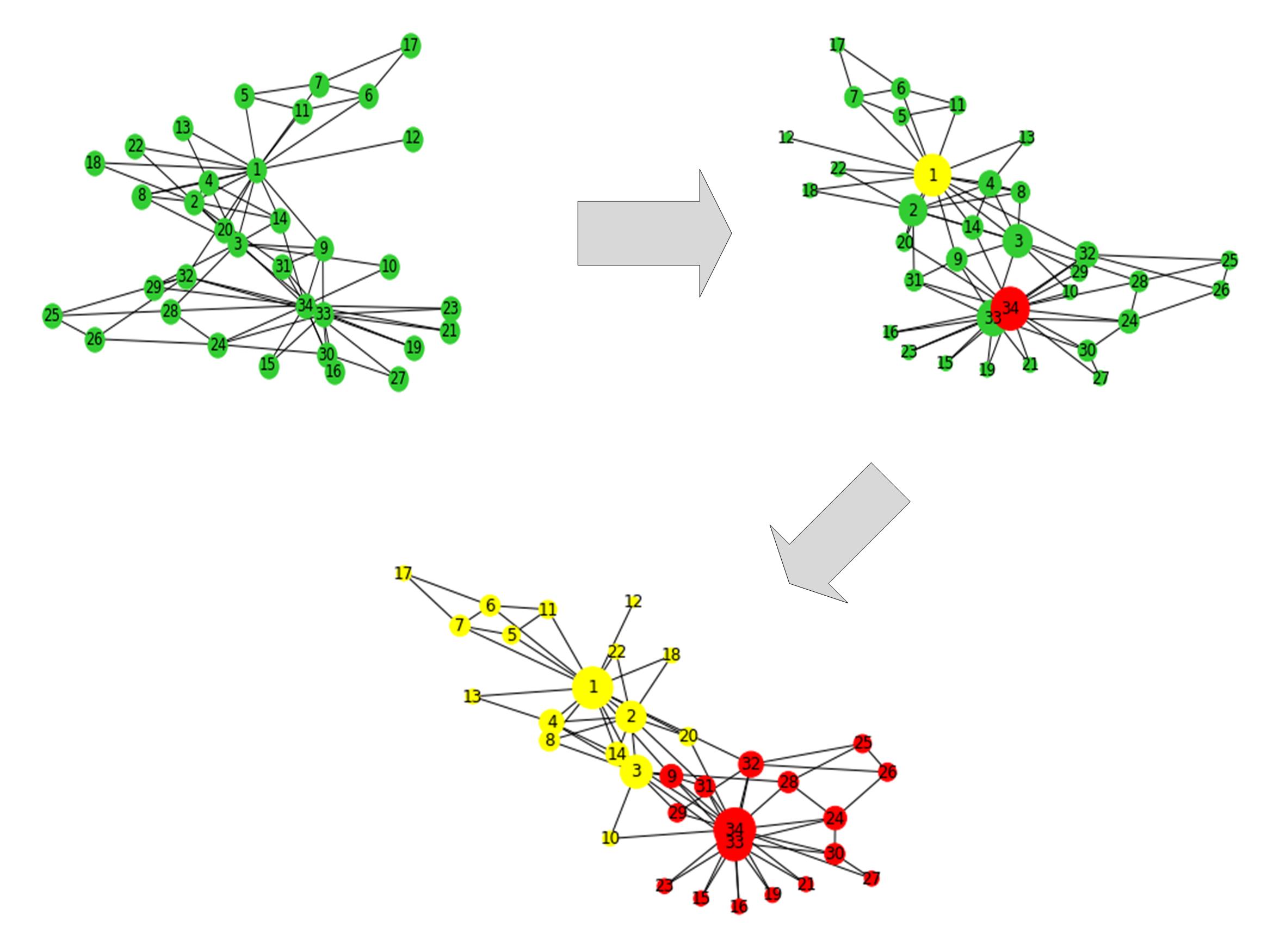}
		\caption{The separation of Karate club 
        network}
		\label{fig:karate}
	\end{minipage}
\end{figure}

Complex network modeling is a powerful tool for quantitatively analyzing social media problems. In a common social network, each node represents an individual, and the links between nodes symbolize the connection between individuals. Take an early network-simulated social activity case as an example: it begins with the emergence of incompatible opinions, leading to polarization in the karate club with the dissemination of standpoints and ultimately resulting in its split \cite{cavallari2017learning} (see \autoref{fig:karate}). A large number of researchers have studied dynamic networks that simulate the formation of information cocoons, focusing on the influence of powerful users and the selective acceptance of information \cite{gu2024modeling}, \cite{flache2014small}, \cite{song2023modeling}. Several investigations focus on information cocoons related to network topology structure, especially community structure \cite{ren2022investigating}, \cite{zhang2022social}. However, the majority of research on community structure merely focuses on single-layer networks, neglecting the algorithmic influence on users. The research related to information cocoon taking both algorithm and network topology structure appears to be absent. To fully explore the topological characteristics influenced by similarity-based recommendation algorithms, we construct a double-layer network that integrates user interactions and user similarity, capable of excavating sub-structures in social relationships and imitating social interactions shaped by data-driven algorithms.

The information cocoons can be analog to sub-structures with two main features. One is that most of the viewpoints in one sub-structure gradually approach consensus with one dominant viewpoint, and a small number of heterogeneous viewpoints lack influence. The other is that sub-structures are relatively closed, which means the sub-structure interacts with the outside world less, and opinions from outside are hard to influence the core members of the group \cite{liu2023career}. Considering the aforementioned characteristics and complex network modeling for information cocoon research, our task is to explore the sub-network within the overall topic information space, community detection.  

As a downstream task of graph representation learning, there emerges a significant volume of approaches for exploring community structures, ranging from the classical spectral clustering  \cite{ng2001spectral}, \cite{newman2013spectral} to deep graph learning \cite{shchur2019overlapping}, 
\cite{pasa2020deep}, \cite{kipf2017semi}, \cite{velivckovic2017graph}, \cite{kipf2016variational}. Nonetheless, efficient unsupervised frameworks for multi-layer community detection remain scarce. Our research proposes two unsupervised frameworks for solving community detection problems in multi-layer networks, leveraging both modularity optimization and graph auto-encoder. The encoder, built upon a two-layer graph convolutional neural network, performs intra-layer feature fusion and inter-layer feature integration using the input adjacency tensor and modularity tensor. For the decoder, we explore two different strategies: the first approximates the true modularity tensor using feature reconstruction tensors from each layer independently, without feature fusion; the second concatenates the feature reconstruction matrices from all layers into a fused representation to approximate the modularity tensor for each layer.

Current social media do not have a mature monitoring mechanism for information cocoons. In order to avoid group polarization, they only take folding or closing debate measures for some abnormal comments. To address this gap, this paper focuses on the information cocoon phenomenon, proposing a more natural exploration and monitoring framework for discovering potential information cocoons as well as an algorithmic intervention strategy for mitigating group polarization. As mentioned in previous research, the main features of information cocoon are external closure and internal homogeneity \cite{liu2023career}. Based on these two characteristics, this paper has designed a community structure-based excavation framework for information cocoons, which can be applied to a social media dataset with user comments. Subsequently, the intervention measure is implemented within those sub-networks identified as trapped in information cocoons. The whole framework of our research is demonstrated as follows.

\begin{figure}[H]
	\centering
	\begin{minipage}[b]{0.85\textwidth}
		\includegraphics[width=\textwidth]{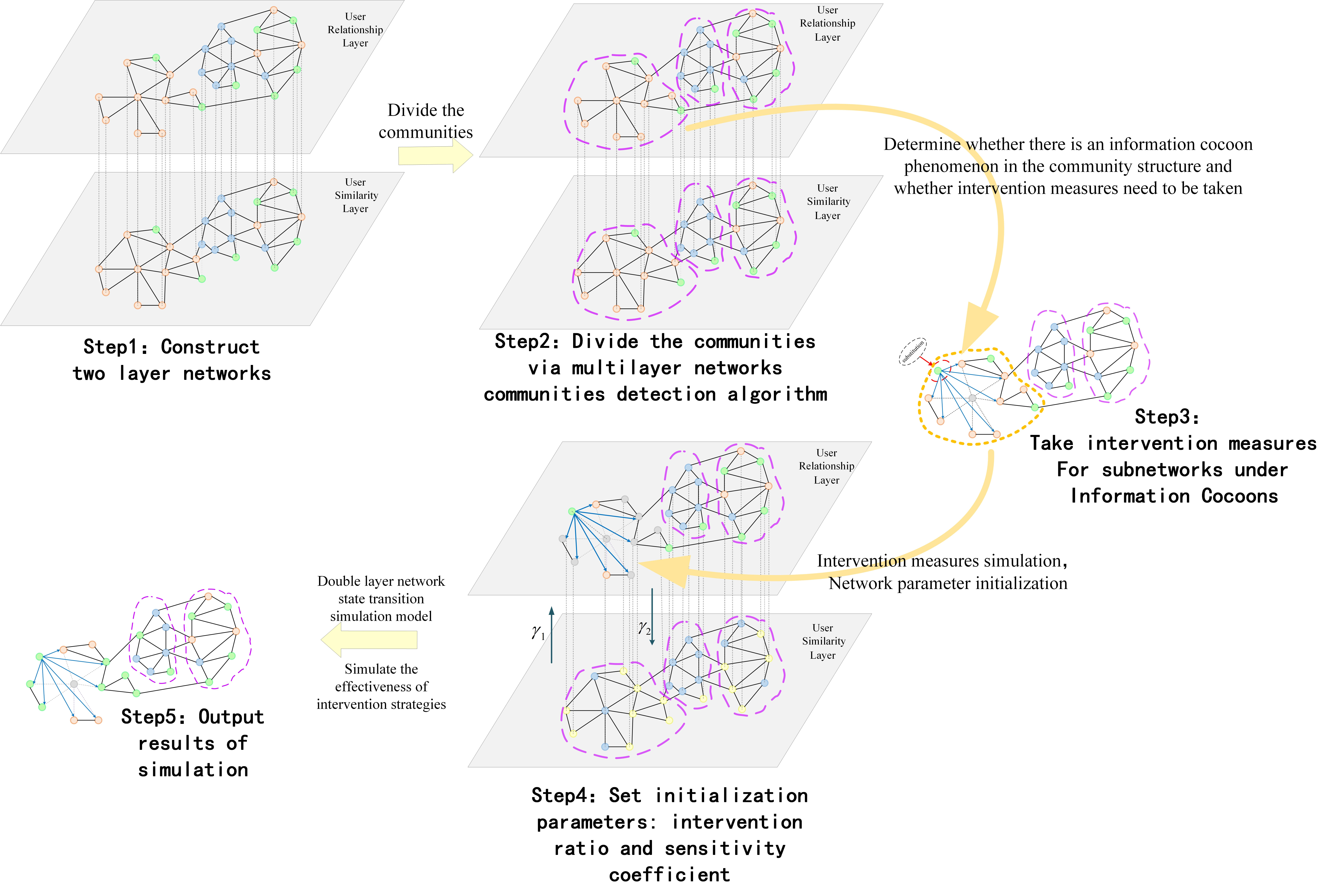}
		\caption{Research framework}
		\label{fig:framework}
	\end{minipage}
\end{figure}

We construct a user relationship-similarity two-layer network based on user interaction correlation and comments feature similarity, as is shown in Step 1 of \autoref{fig:framework}. Next, segment the network through our proposed multi-layer network community detection algorithm in Step 2 and impose intervention measures on some of the sub-networks trapped in information cocoons. To be more precise, the highly influential nodes in sub-networks are substituted by other heterogeneous perspectives. According to the previous association, those heterogeneous perspectives are recommended to other users, as illustrated in Step 3. In the remaining procedures, we simulate the propagation process by our Markov transition model.

The main contributions of this paper are summarized as follows.

\begin{itemize}
	\item We study the phenomenon of information cocoon in the form of quantitative models. We consider the excavation of information cocoons as a dual-objective optimization problem based on graph data due to the homogeneity of viewpoints among users within the same information cocoons. This information cocoon excavation method provides a monitoring and warning mechanism.
	
	\item We propose a novel method for multi-layer network community detection tasks based on graph representation learning. This paper promotes the graph auto-encoder (GAE) method in single-layer networks into multi-layer networks, proposing two modularity tensor reconstruction algorithms based on GAE. The first one is a mixed graph embedding-based modularity tensor reconstruction method (MGE-MTR). This method concatenates each feature matrix from the encoder into a single feature matrix, reconstructing each slice of the real modularity tensor during decoder process. The second one is the independent graph embedding-based modularity tensor reconstruction method (IGE-MTR), which utilizes each slice of graph representation tensor to reconstruct the modularity tensor.
	
	\item We provide a robust intervention measure that can be flexibly applied to algorithms. In the meantime, the simulation experiments verify our proposed intervention measures could mitigate the polarization of groups to a certain extent.
	
	\item The utilization of social network data is distinctive from other references. Most previous research primarily focuses on  specific types of user connections \cite{pastor2001epidemic}. However, we introduce a similarity network generated by probability constructed from the cosine value of users' features: \begin{align*}
		\mathbb{P}\left(a_{ij}=1|\overrightarrow{\bm{z}}_{i},\overrightarrow{\bm{z}}_{ij} \right)=\cos\left\langle\overrightarrow{\bm{z}}_{i},\overrightarrow{\bm{z}}_{j} \right\rangle =\frac{\overrightarrow{\bm{z}}_{i}\cdot\overrightarrow{\bm{z}}_{j}}{\left|\overrightarrow{\bm{z}}_{i}\right|\cdot\left|\overrightarrow{\bm{z}}_{j}\right|},  
	\end{align*}
	where $a_{ij}$ is the $i$-th row $j$-th column adjacent matrix, whose value equals 1 representing there exists a connection between the $i$-th and $j$-th nodes. And $\overrightarrow{\bm{z}}_{i}$ is the feature vector of the $i$-th node.
	It has significant meaning for excavating potential connections when adding the similarity layer. 
\end{itemize}

The remainder of the paper is organized as follows. \autoref{section: Related Work} reviews the related work. 
In \autoref{section: Preliminaries and Main Result}, we illustrate a basic set of multi-layer network community detection problems and offer theoretical insurance.  \autoref{section: Proposed Method} presents our novel methods proposed in this paper, including two unsupervised frameworks for addressing community detection problems in multi-layer networks leveraging modularity optimization and graph auto-encoder, as well as a user influence-based intervention strategy and double-layer intervention effect simulation system. 
Compared with existing multi-layer network unsupervised community detection algorithms, these two proposed algorithms are implemented into public Cora, Citeseer, and synthetic datasets in \autoref{section: Numerical Analysis}.
\autoref{section: Real Data Analysis} mainly focuses on real-world data analysis. Firstly, we construct double-layer networks through the Weibo dataset and implement our algorithm into this network to explore the latent information cocoons. Next, we adopt our proposed intervention strategy and apply a simulation system model to imitate standpoint dissemination in a double-layer network constructed by real-world data. In \autoref{section: Discusion}, we make a sensitivity analysis for some of the parameters in our simulation system. Meanwhile, we come up with the corresponding conclusion about the consequences of group polarization and information cocoons. The conclusion and further discussion can be found in \autoref{section: Conclusion}.

\section{Related Work} \label{section: Related Work}

\subsection{Information Cocoon}
Most previous research focuses on qualitative research, which is conducted in the form of questionnaires. Ren et al.\cite{ren2022investigating} investigated several short-form video users by questionnaire and came up with the main factor of forcing or hindering information cocoon generation. More concretely, the subjective preference of users and stable interest recommendations could accelerate the formation of information cocoons, while receiving heterogeneous information and random recommendations could impede the generation of information cocoons. Tracking the behavior of users, Li et al. \cite{li2022exploratory} proposed solutions to information cocoons from a macro and systematic perspective-community structure. Hou et al. \cite{hou2023information} simulated the information propagation in social networks with the effect of recommendation systems. This led to the conclusion that similarity-based recommendation systems could intensify the formation of information cocoons. Intriguing by these results, we begin with a multi-layer network structure and design a quantity method by excavating the sub-structure in social platform data for analyzing information cocoons.

\subsection{Social Network Modeling}
The early development of community-based social network models primarily relied on dissemination principles, most notably those derived from the discrete SIR (Susceptible-Infected-Recovered) model introduced by Kermack and McKendrick in 1927 \cite{brauer2005kermack}. A link between two nodes is based on whether a direct connection exists between the disseminator and recipients. As the dissemination properties are similar between public sentiment and infection, the information propagation process can be simulated by this model. Based on the basic infection transmission model, Daley and Kendal proposed the DK rumor dissemination model \cite{daley1964epidemics}. This model endowed nodes with three statuses for simulating rumor dissemination in networks. Like rumors and sentiments, a node can pass its disease to its neighbors. Such a model is gradually developed based on complex network theory and network topology structure for the research of individual association and dissemination behavior \cite{pastor2001epidemic}. Recently, in most social network research, the connections are established through real contact between entities, such as interactions, relationships, friends, and colleagues \cite{al2019improving}. This network model is restricted in studying diverse propagation patterns, as it solely accounts for a single-layer structure, neglecting potential inter-layer coupling information.

Most datasets appear to have a strong connection between entities and contain obvious network topology structures, such as global trade networks and protein molecular data. Unlike those datasets, social network data contain much latent information apart from interaction relationships.

With the diversity of propagation ways and channels, the dissemination of information contains two or more mutual effects, meaning the single-layer network cannot entirely reflect the information propagation process. Under different scenes, researchers established multilayer networks. Based on the infectious model (SIR), Yagan et al. \cite{yagan2013conjoining} established society-physical double-layer information diffusion networks and further explored the interaction effect of information dissemination in double-layer networks. Based on two online social platform-twitter and Facebook, Magnani et al. \cite{magnani2011ml} established a two-layer network and proposed a multilayer graph representation learning model, which reflects the interaction between different platforms. Considering the single social media platform Facebook, Xiong et al. \cite{xiong2018emotional} extracted four kinds of contact relation: following, forwarding, mentioning, and replying, constructing multilayer networks, and analyzing the sentiment dissemination in social networks. These networks are constructed by entities' real contacts, such as following, mentioning, and replying, which seems sparse. Ioannidis et al. \cite {ioannidis2019recurrent} established a two-layer network utilizing Cora and Citeseer datasets through real citation relations and $k$-th nearest neighbors of features. The training results of graph recurrent neural networks show that such two-layer networks containing feature connections can reveal multiple structures \cite{ioannidis2019recurrent}. 

\subsection{Community Detection}
Network structure data can be divided into several disjoint vertices subsets, ensuring the connections within sub-networks are dense, and the connections between sub-networks are sparse. Excavate the potential sub-structure of networks, which have extensive applications in many areas. For instance, the co-research areas can be discovered through segmenting cooperation networks; the marketing strategies can be designed by exploring similar users, the protein structure network can reveal the interconnection between molecules, and partition social networks can be used to track rumor dissemination. 

They have developed many algorithms for single-layer networks, such as graph segmentation and clustering. Graph segmentation mainly contains sub-graph partition and modularity optimization, two methods aiming at converting community partition into optimization problems. Ng et al. \cite{ng2001spectral} proposed a sub-graph partition algorithm by setting the Laplacian matrix as the objective function. Newman et al. \cite{newman2004finding} promoted the graph partition method through a modularity matrix. After that, advanced algorithms based on modularity optimization sprung out. Modularity-based community detection algorithms applied in the directed and weighted graphs were proposed in 2007 \cite{arenas2007size}. Louvain algorithm improved from modularity optimization was proposed in 2008, making it applicable to large-scale networks \cite{blondel2008fast}. The overlapping community segmentation problems in undirected and unweighted graphs were solved by Shen et al. \cite{shen2009detect} in 2009. Graph clustering approaches achieve entity labels through graph representation feature vectors, mainly containing spectral cluster \cite{newman2013spectral} and stochastic block model-based statistical inference method \cite{lee2019review}. With the development of deep learning, some graph representation learning methods achieved fantastic effects in the downstream task community detection, such as recurrent graph neural network \cite{pasa2020deep}, graph convolutional neural network \cite{kipf2017semi}, graph attention neural networks \cite{velivckovic2017graph}, graph auto-encoder \cite{kipf2016variational} and so on. All the above algorithms are non-linear supervised or semi-supervised learning methods to obtain low-rank graph representations, which require training samples with priori labels. Whereas most real-world data lack previous information. Some works attempted to integrate label sampling phrase and graph representation learning phrase into an unsupervised two-stage framework \cite{verma2023efficient}, \cite{chen2019exploiting}, \cite{rostami2022novel}. 
Leveraging modularity and topological information, Choong et al. \cite{choong2018learning} proposed a variational graph auto-encoder reconstruction algorithm, which could be applied to unsupervised community detection tasks, without requiring priori label information.

Applying the single-layer network community detection algorithms directly into multi-layer structures is challenging. Research on multi-layer networks can be broadly divided into two categories: one based on matrix decomposition and clustering and the other on deep learning–based graph representation methods. The traditional matrix decomposition and cluster method tries to fuse the several layers of networks by mapping the multi-layer networks into the single-layer counterparts or integrating the labeled vertices across multiple layers of the networks. Apart from the above methods, deep graph representation learning has attracted much attention, most of which integrate graph embeddings with representation vectors clustering into two stages. Bahadori et al. \cite{bahadori2021improved} designed a fusion strategy of local random walking for multi-layer networks. Song and Thiagarajan \cite{song2019improved} proposed a deep random walking model combining the traditional random walking method and deep learning. The node could randomly walk in coupling edges, thus achieving low-rank representation. Naderipour et al. \cite{naderipour2020type} proposed a possible c-means cluster model that could utilize topology structure and similarity features to divide the overlapping communities in large-scale networks. Mansoureh et al. \cite{mansoureh2022multilayer} considered a newly defined degree and proposed a multi-layer general type-2 fuzzy community detection model. Paul \cite{paul2021null} constructed multi-layer network confusion modularity degree and expected modularity degree to find the optimal community labels. Ioannidis et al. \cite{ioannidis2019recurrent} proposed a semi-supervised graph recurrent neural network framework for multi-layer networks, achieving better performance in node classification tasks.

\subsection{Reconstruction of Modularity}
Maximization of modularity for community partition was proposed in 2006 \cite{newman2006modularity}, which is expected to include as many edges as possible in the same community. Let the indicator variable $Z(i,j)\in\{0,1\}$. When node $i$ and node $j$ belong to the same community, $Z(i,j)$ equals $1$. Otherwise, its value is $0$. $k_i$ means the degree of node $i$. $m=\frac{1}{2}\sum_{i}k_i$ represents the total number of edges in the whole network. The definition of modularity is:
\begin{align*}
	Q=\frac{1}{2m}\sum\limits_{i,j}\left(a_{ij}-\frac{k_ik_j}{2m} \right)Z(i,j). 
\end{align*}
The modularity matrix $B=[b_{ij}]_{N\times N}\in \mathbb{R}^{N\times N}$ is constructed in \cite{bhowmick2024dgcluster}, where $b_{ij}=a_{ij}-\frac{k_ik_j}{2m}$. Let community feature matrix $Z=[Z_1,Z_2,\cdots,Z_K]\in\{0,1\}^{N\times K}$. The modularity can be represented as:
\begin{align*}
	Q=\frac{1}{2m}\sum\limits_{i=1}^{K}Z_i^TBZ_i=\frac{1}{2m}Tr\left(Z^TBZ \right). 
\end{align*}
The optimization problem can be expressed as: 
\begin{align*}
	\max\limits_{\substack{Z\in\{0,1\}^{N\times K}\\ \sum\limits_{i=1}^{K}Z_i=\mathbf{1}_N}}\frac{1}{2m}\sum\limits_{i=1}^{K}Z_i^TBZ_i.
\end{align*}
It seems to be an NP hard problem. Then, we can write down the relaxation version:
\begin{align*}
	\max\limits_{Tr(Z^TZ)=N}Tr\left(Z^TBZ \right).
\end{align*}
The optimal value of the above problem is equal to the largest eigenvalue of modularity matrix B. Based on Matrix Reconstruction Theorem \cite{eckart1936approximation}, for any modularity matrix of order N, there exists an approximation matrix $\hat{B}$ of order $r$ (where $r<N$) that closely estimates it. Consequently, the modularity maximization problem can be viewed as a task of finding the optimal low-rank approximation.
\begin{align*}
	\min\limits_{\substack{\hat{B}\in\mathbb{R}^{N\times N}\\ r(\hat{B})\le r}}\left\|B-\hat{B}\right\|_F\Leftrightarrow\min\limits_{\substack{\hat{B}\in\mathbb{R}^{N\times N}\\ r(\hat{B})\le r}}\sqrt{\lambda_{r+1}^2+\lambda_{r+2}^2+\cdots+\lambda_{N}^2}.
\end{align*}
This minimization problem can be solved by nonlinear methods such as neural networks \cite{bhowmick2024dgcluster}, \cite{salha2022modularity}, \cite{murata2018modularity}.

\section{Preliminaries}\label{section: Preliminaries and Main Result}
\subsection{Community Detection Problem in Multi-layer Networks}\label{subsection: Community Detection Problem in Multi-layer Networks}
For multi-layer networks $\mathcal{G}=(\mathcal{V},\mathcal{E})$, $\mathcal{V}$ represents nodes in $L$-th layer and $\mathcal{E}=\{E^1,\cdots,E^L\}$ is the edge set in $L$ layers. The adjacent tensor is $\mathcal{A}=[A^1,\cdots,A^L]$, where $A^l=[a_{ij}^l]_{N\times N}$ is adjacent matrix of the $l$-th layer. The modularity matrix of $l$-th layer is $B^{(l)}=a_{ij}^l-\frac{k_i^lk_j^l}{m^l}$ where $k_i^l$ is the degree of the node $i$ of $l$-th layer. A community detection task in multi-layer networks aims to discover node representation vectors to maximize the modularity value of both tensors. 
\begin{equation}
	\begin{split}
		\max\quad Q(Z)=&\left[Tr(Z^TB^{(1)}Z),Tr(Z^TB^{(2)}Z),\cdots,Tr(Z^TB^{(L)}Z)\right],\\
		s.t.\quad &Tr(Z^TZ)=N.
	\end{split}\label{main problem}
\end{equation}

We naturally arrive at using a linear weighting method to transform the multi-objective optimization problem into a single-objective maximization problem.
\begin{equation}
	\begin{split}
		\max\quad \alpha_1Tr(Z^TB^{(1)}Z)+&\alpha_2 Tr(Z^TB^{(2)}Z)+\cdots+\alpha_L Tr(Z^TB^{(L)}Z),\\
		s.t.\quad &Tr(Z^TZ)=N,\\
		&\alpha_1+\cdots+\alpha_L=1,\\
		&\alpha_1>0,\cdots,\alpha_L>0.
	\end{split}\label{single obective maximization}
\end{equation}

\subsection{Theoretical Analysis}\label{subsection:Theoretical Analysis}
\begin{theorem}\label{optimal low rank reconstruction}
	The maximization problem \eqref{single obective maximization} is equivalence to finding the low rank approximation of matrix $\Theta=\left[\begin{matrix}
		B^{(1)}&\cdots&O\\
		\vdots&\ddots&\vdots\\
		O&\cdots&B^{(L)}
	\end{matrix} \right] .$ The low rank approximation matrix $\hat{\Theta}$ can be formulated as the dot product of feature matrix $\Phi=\left[\begin{matrix}
	    \Phi_1&\cdots&O\\
            \vdots&\ddots&\vdots\\
            O&\cdots&\Phi_L
	\end{matrix}\right]$, where $\Phi_l\in\mathbb{R}^{N\times r_l}$ is the reconstruction feature matrix of modularity matrix $B^{(l)}$, namely $\hat{B}^{(l)}=\Phi_l\Phi_l^T.$
\end{theorem}

\autoref{optimal low rank reconstruction} provides the theoretical guarantee for our independent graph embedding-based modularity tensor reconstruction approach (IGE-MTR). The $NL\times\tau$ dimension representation matrix means that the concatenation of the reconstruction features from each layer of the modularity tensor $B^{(l)}$ is equivalent to the reconstruction of the high-dimensional block matrix $\Theta$. This enables the solution of \eqref{single obective maximization} by reconstructing each layer of modularity tensor $B^{(l)}$, followed by concatenating all the reconstructed feature matrices.

\begin{theorem}\label{optimal low rank reconstruction2}
	The maximization problem \eqref{single obective maximization} is equivalence to finding the low rank approximation of matrix $M=\alpha_1B^{(1)}+\alpha_2B^{(2)}+\cdots+\alpha_LB^{(L)}.$
    The low-rank approximation matrix $\hat{M}$ can be formulated as the dot product of the representation feature matrix $\Gamma=\left[\begin{matrix}
    \sqrt{\alpha_1}\Phi_1&\sqrt{\alpha_2}\Phi_2&\cdots&\sqrt{\alpha_L}\Phi_L
    \end{matrix}\right]\in\mathbb{R}^{N\times\tau}$, that is $\hat{M}=\Gamma\Gamma^T$.
\end{theorem}

\autoref{optimal low rank reconstruction2} demonstrates that the single feature matrix $\Gamma\in\mathbb{R}^{N\times\tau}$ is capable of reconstructing each layer of the modularity tensor, thereby supporting the community detection algorithm based on fused graph representation features. Since 
\begin{equation}
	\begin{split}
		\left\|M-\hat{M}\right\|_F^2=&\left\|\alpha_1B^{(1)}+\alpha_2B^{(2)}+\cdots+\alpha_LB^{(L)}-\hat{M}\right\|_F^2\\
		\le&\left\|\alpha_1B^{(1)}-\alpha_1\hat{M}\right\|_F^2+\left\|\alpha_2B^{(2)}-\alpha_2\hat{M}\right\|_F^2+\cdots+\left\|\alpha_LB^{(L)}-\alpha_L\hat{M}\right\|_F^2\\
		\le&\left\|B^{(1)}-\hat{M}\right\|_F^2+\left\|B^{(2)}-\hat{M}\right\|_F^2+\cdots+\left\|B^{(L)}-\hat{M}\right\|_F^2,
	\end{split}\label{MGRF}
\end{equation}
minimizing the distance $\left\|M-\hat{M}\right\|_F$ can be relaxed into minimizing the sum of each layer reconstruction distance $\left\|B^{(1)}-\hat{M}\right\|_F+\left\|B^{(2)}-\hat{M}\right\|_F+\cdots+\left\|B^{(L)}-\hat{M}\right\|_F$.

\begin{proof}[Proof of \autoref{optimal low rank reconstruction}]
The problem \eqref{single obective maximization} is equivalence to:
\begin{equation}
	\begin{split}
		\max\quad&Tr\left( \left[\begin{matrix}
			Z^T,\cdots,Z^T
		\end{matrix} \right] \left[\begin{matrix}
			\alpha_1B^{(1)}&\cdots&O\\
			\vdots&\ddots&\vdots\\
			O&\cdots&\alpha_LB^{(L)}
		\end{matrix} \right] \left[\begin{matrix}
			Z^T\\ \vdots\\Z^T
		\end{matrix} \right]\right) , \\
		s.t.\quad& Tr(Z^TZ)=N,\\
		&\alpha_1+\cdots+\alpha_L=1,\\
		&\alpha_1>0,\alpha_2>0,\cdots,\alpha_L>0.
	\end{split}\label{version 1}
\end{equation}
The problem \eqref{version 1} can also be written as:
\begin{equation}
	\begin{split}
		\max\quad&Tr\left( \left[\begin{matrix}
			\sqrt{\alpha_1}Z^T,\cdots,\sqrt{\alpha_L}Z^T
		\end{matrix} \right] \left[\begin{matrix}
			B^{(1)}&\cdots&O\\
			\vdots&\ddots&\vdots\\
			O&\cdots&B^{(L)}
		\end{matrix} \right] \left[\begin{matrix}
			\sqrt{\alpha_1}Z^T\\ \vdots\\\sqrt{\alpha_L}Z^T
		\end{matrix} \right]\right) , \\
		s.t.\quad& Tr(\alpha_1Z^TZ)+Tr(\alpha_2Z^TZ)+\cdots+Tr(\alpha_LZ^TZ)=N,\\
		&\alpha_1>0,\alpha_2>0,\cdots,\alpha_L>0.
	\end{split}\label{version 2}
\end{equation}
Let $\Theta=\left[\begin{matrix}
	B^{(1)}&\cdots&O\\
	\vdots&\ddots&\vdots\\
	O&\cdots&B^{(L)}
\end{matrix}\right]$ and $U=\left[\begin{matrix}
	\sqrt{\alpha_1}Z\\ \vdots\\ \sqrt{\alpha_L}Z
\end{matrix}\right]$, \eqref{version 1} can be formulated as:
\begin{align*}
	&\arg\max\limits_{\substack{Tr(Z^TZ)=N\\ \alpha_1+\cdots+\alpha_L=1\\ \alpha_1>0,\cdots,\alpha_L>0}} \alpha_1Q^{(1)}+\alpha_2Q^{(2)}+\cdots+\alpha_LQ^{(L)}\\
	=&\arg\max\limits_{Tr(U^TU)=N}Tr(U^T\Theta U).
\end{align*}
Through eigenvalue decomposation, $\Theta$ can be written as the multiple of block matrix:
\begin{align*}
	\Theta=\left[\begin{matrix}
		H^{(1)}&P^{(1)}&\cdots&O&O\\
		\vdots&\vdots&\ddots&\vdots&\vdots\\
		O&O&\cdots&H^{(L)}&P^{(L)}
	\end{matrix}\right]
	\left[\begin{matrix}
		\Lambda_{r_1}^{(1)}&O&&&&\\
		O&\Lambda_{N-r_1}^{(1)}&&&&\\
		&&\ddots&&\\
		&&&\Lambda_{r_L}^{(L)}&O\\
		&&&O&\Lambda_{N-r_L}^{(L)}
	\end{matrix}\right]
	\left[\begin{matrix}
		H^{(1)T}&\cdots&O\\
		P^{(1)T}&\cdots&O\\
		\vdots&\ddots&\vdots\\
		O&\cdots&H^{(L)T}\\
		O&\cdots&H^{(L)T}
	\end{matrix}\right].
\end{align*}
Intriguing by \autoref{low rank approximation}, we can find the $r_1+\cdots+r_L$-order low-rank approximation
\begin{align*}
	\hat{\Theta}=\left[\begin{matrix}
		H^{(1)}&\cdots&O\\
		\vdots&\ddots&\vdots\\
		O&\cdots&H^{(L)}
	\end{matrix}\right]
	\left[\begin{matrix}
		\Lambda_{r_1}^{(1)}&\cdots&O\\
		\vdots&\ddots&\vdots\\
		O&\cdots&\Lambda_{r_L}^{(L)}
	\end{matrix}\right]
	\left[\begin{matrix}
		H^{(1)T}&\cdots&O\\
		\vdots&\ddots&\vdots\\
		O&\cdots&H^{(L)T}
	\end{matrix}\right].
\end{align*}
Denote \begin{align*}
	\Lambda_{\tau}=\left[\begin{matrix}
		\Lambda_{r_1}^{(1)}&\cdots&O\\
		\vdots&\ddots&\vdots\\
		O&\cdots&\Lambda_{r_L}^{(L)}
	\end{matrix}\right],
\end{align*}
where $\tau=r_1+\cdots+r_L$ and $r_l=rank\left(\Lambda_{r_l}^{(l)} \right). $ Since the eigenvalue diagonal matrix $\Lambda_\tau$ is composed of positive eigenvalue, it can be represented as $\Lambda_\tau=\Sigma_\tau\Sigma_\tau$, where $\Sigma_\tau$ can be formulated as a block matrix:\\
\begin{align*}
    \Sigma_\tau=\left[\begin{matrix}
        \Sigma_{r_1}^{(1)}&\cdots&O\\
        \vdots&\ddots&\vdots\\
        O&\cdots&\Sigma_{r_L}^{(L)}
    \end{matrix}\right].
\end{align*}

Let $\Phi=\left[\begin{matrix}
	H^{(1)}&\cdots&O\\
	\vdots&\ddots&\vdots\\
	O&\cdots&H^{(L)}
\end{matrix}\right]\Sigma_\tau\in\mathbb{R}^{NL\times\tau}$ , thus $\hat{\Theta}=\Phi\Phi^T$ and \begin{align*}
    \Phi=\left[\begin{matrix}
	H^{(1)}\Sigma_{r_1}^{(1)}&\cdots&O\\
	\vdots&\ddots&\vdots\\
	O&\cdots&H^{(L)}\Sigma_{r_L}^{(L)}
\end{matrix}\right]=\left[\begin{matrix}
	\Phi_1&\cdots&O\\
	\vdots&\ddots&\vdots\\
	O&\cdots&\Phi_L
\end{matrix}\right],
\end{align*} where $\Phi_l=H^{(l)}\Sigma_{r_l}^{(l)}\in\mathbb{R}^{N\times r_l}$ is the representation matrix of $B^{(l)}$ since $\hat{B}^{(l)}=H^{(l)}\Sigma_{r_l}^{(l)}\Sigma_{r_l}^{(l)T}H^{(l)T}\\=H^{(l)}\Lambda_{r_l}^{(l)}H^{(l)T}$. The multi-objective optimization problem \eqref{single obective maximization} can be converted into low-rank approximation of $\Theta$:
\begin{align*}
	&\arg\max\limits_{\substack{Tr(Z^TZ)=1\\ \alpha_1+\cdots+\alpha_L=1\\ \alpha_1>0,\cdots,\alpha_L>0}} \alpha_1Q^{(1)}+\alpha_2Q^{(2)}+\cdots+\alpha_LQ^{(L)}\\
	=&\arg\min\limits_{\substack{\hat{\Theta}\in\mathbb{R}^{NL\times NL}\\
			r(\hat{\Theta})\le \tau}}\left\|\Theta-\hat{\Theta}\right\|_{F}\\
	=&\arg\min\limits_{\substack{\Phi\in\mathbb{R}^{NL\times\tau}\\
			r(\hat{\Phi})\le \tau}}\left\|\Theta-\Phi\Phi^T\right\|_{F}.
\end{align*}
The proof of \autoref{optimal low rank reconstruction} is then completed.
\end{proof}

\begin{proof}[Proof of \autoref{optimal low rank reconstruction2}]
From \autoref{optimal low rank reconstruction}, we know that the matrix $\Theta$ can be approximated by $\hat{\Theta}=\Phi\Phi^T$.
Let 
\begin{align*}
	M=\alpha_1B^{(1)}+\alpha_2B^{(2)}+\cdots+\alpha_LB^{(L)}=\left[\begin{matrix}
		\sqrt{\alpha_1}\bm{I}_N&\cdots&\sqrt{\alpha_L}\bm{I}_N
	\end{matrix}\right]\Theta
	\left[\begin{matrix}
		\sqrt{\alpha_1}\bm{I}_N\\
		\vdots\\
		\sqrt{\alpha_L}\bm{I}_N
	\end{matrix}\right]\in\mathbb{R}^{N\times N}.
\end{align*}
The $\tau$ order rank approximation of M is:
\begin{align*}
	\hat{M}=&\left[\begin{matrix}
		\sqrt{\alpha_1}\bm{I}_N&\cdots&\sqrt{\alpha_L}\bm{I}_N
	\end{matrix}\right]\hat{\Theta}
	\left[\begin{matrix}
		\sqrt{\alpha_1}\bm{I}_N\\
		\vdots\\
		\sqrt{\alpha_L}\bm{I}_N
	\end{matrix}\right]\\
	=&\left[\begin{matrix}
		\sqrt{\alpha_1}\bm{I}_N&\cdots&\sqrt{\alpha_L}\bm{I}_N
	\end{matrix}\right]
	\left[\begin{matrix}
		H^{(1)}&\cdots&O\\
		\vdots&\ddots&\vdots\\
		O&\cdots&H^{(L)}
	\end{matrix}\right]
	\Lambda_{\tau}
	\left[\begin{matrix}
		H^{(1)T}&\cdots&O\\
		\vdots&\ddots&\vdots\\
		O&\cdots&H^{(L)T}
	\end{matrix}\right]
	\left[\begin{matrix}
		\sqrt{\alpha_1}\bm{I}_N\\
		\vdots\\
		\sqrt{\alpha_L}\bm{I}_N
	\end{matrix}\right]\\
	=&\left[\begin{matrix}
		\sqrt{\alpha_1}H^{(1)}&\cdots&\sqrt{\alpha_L}H^{(L)}
	\end{matrix}\right]\Lambda_{\tau}
	\left[\begin{matrix}
		\sqrt{\alpha_1}H^{(1)T}\\
		\vdots\\
		\sqrt{\alpha_L}H^{(L)T}
	\end{matrix}\right].
\end{align*}
Let $\tilde{H}^{(l)}=\sqrt{\alpha_l}H^{(l)}$, then \begin{align*}
    \hat{M}=\left[\begin{matrix}
	\tilde{H}^{(1)}&\cdots&\tilde{H}^{(L)}
\end{matrix}\right]\Lambda_{\tau}
\left[\begin{matrix}
	\tilde{H}^{(1)T}\\
	\vdots\\
	\tilde{H}^{(L)T}
\end{matrix}\right],
\end{align*} where $\tau=r_1+\cdots+r_L$ and $r_l=rank\left(\Lambda_{r_l}^{(l)} \right). $ Since the eigenvalue diagonal matrix $\Lambda_\tau$ is composed of positive eigenvalue, it can be represented as $\Lambda_\tau=\Sigma_\tau\Sigma_\tau$.

Let $\Gamma=\left[\begin{matrix}
	\tilde{H}^{(1)}&\cdots&\tilde{H}^{(L)}
\end{matrix}\right]\Sigma_\tau\in\mathbb{R}^{N\times\tau}$, thus $\hat{M}=\Gamma\Gamma^T$. And the representation matrix $\Gamma$ can be formulated by $\Phi_l$: $\Gamma=\left[\begin{matrix}
\sqrt{\alpha_1}\Phi_1&\sqrt{\alpha_2}\Phi_2&\cdots&\sqrt{\alpha_L}\Phi_L
    \end{matrix}\right]$. The multi-objective optimization problem \eqref{single obective maximization} can be converted into low-rank approximation of $\Theta$ or $M$:
\begin{align*}
	&\arg\max\limits_{\substack{Tr(Z^TZ)=N\\ \alpha_1+\cdots+\alpha_L=1\\ \alpha_1>0,\cdots,\alpha_L>0}} \alpha_1Q^{(1)}+\alpha_2Q^{(2)}+\cdots+\alpha_LQ^{(L)}\\
	=&\arg\min\limits_{\substack{\hat{M}\in\mathbb{R}^{N\times N}\\
			r(\hat{M})\le \tau}}\left\|M-\hat{M}\right\|_{F}\\
	=&
    \arg\min\limits_{\substack{\Gamma\in\mathbb{R}^{N\times\tau}\\
			r(\Gamma)\le \tau}}\left\|M-\Gamma\Gamma^T\right\|_{F}.
\end{align*}
The proof of \autoref{optimal low rank reconstruction2} is then completed.
\end{proof}

\section{Proposed Methods} \label{section: Proposed Method}
\subsection{Independent Graph Embedding Algorithm for Multi-layer Network Community Detection}\label{subsection:Identity Graph Tensor Representation Features}
Graph auto-encoder (GAE) was applied to reconstruct the modularity matrix. 
In this part, we develop a novel GAE-based multi-layer network community detection algorithm.
The graph representation vectors containing community information obtained from the encoder are subsequently applied to reconstruct the modularity tensor.

Based on the modularity tensor optimization problem \eqref{main problem}, we first develop an independent graph embedding-based  modularity tensor reconstruction algorithm (IGE-MTR). Input modularity tensor $B=\left[B^{(1)},\cdots,B^{(L)}\right]$ composed of $L$ layers symmetry modularity matrix and multi-layer adjacent tensor $A=\left[A^1,\cdots,A^L\right]$ into  $L$ layers graph convolutional encoder. The $m$-th graph convolution is formulated as $GCN_m^1\left(A^m,B^{(m)} \right)=\sigma\left(A^mB^{(m)}W_1^m \right),\quad 1\le m\le L $, where $\sigma(\cdot)$ is activation function. Since modularity matrices are always sparse, we choose $\tanh(\cdot)$ as the activation function. 

\begin{figure}[H]
	\centering
	\begin{minipage}[b]{0.9\textwidth}
		\includegraphics[width=\textwidth]{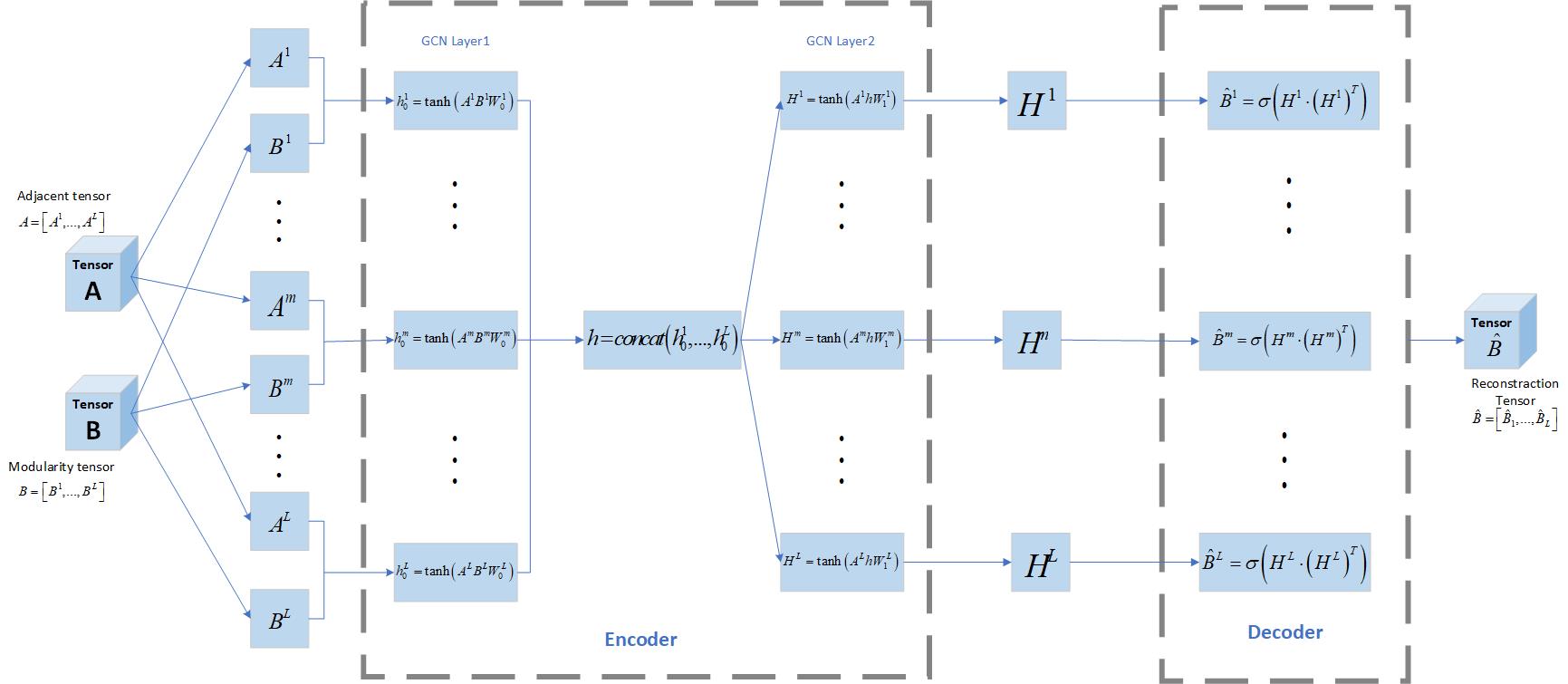}
		\caption{Multi-layer network community detection (IGE-MTR)}
		\label{fig:Identity Graph Representation Features}
	\end{minipage}
\end{figure}

The encoder consists of two sets of graph convolutional layers, each of which comprises $L$ graph convolution layers and operates feature aggregation and transformation for input matrices $B^{(m)}$ and $A^m$, and outputs L graph representation matrices. A feature fusion operation is performed before being fed into the second graph convolutional module. The outputs of the first graph convolutional module $h_0^1=\tanh(A^1B^{(1)}W_0^1),\cdots,h_0^L=\tanh(A^LB^{(L)}W_0^L)$ are concatenated as $h=concat(h_0^1,\cdots,h_0^L)$. Then input the aggregation feature $h$ into the second set of feature feature fusion operation $GCN_m^2\left(A^m, h \right)=\sigma\left(A^mhW_1^m \right) $. \autoref{fig:Identity Graph Representation Features} exhibits our proposed IGE-MTR algorithm.

 The modularity reconstruction effect is evaluated by the sum of the F-norm. From \autoref{optimal low rank reconstruction}
\begin{align*}
    \left\|\Theta-\hat{\Theta}\right\|_F^2=&Tr\left(\sum\limits_{l=1}^{L}\left(B^{(l)}-\hat{B}^{(l)}\right)\left(B^{(l)}-\hat{B}^{(l)}\right)^T\right)\\
    =&\sum\limits_{l=1}^{L}Tr\left(\left(B^{(l)}-\hat{B}^{(l)}\right)\left(B^{(l)}-\hat{B}^{(l)}\right)^T\right)\\
    =&\sum\limits_{l=1}^{L}\left\|B-\hat{B}\right\|_F^2,
\end{align*} we set the loss function $L\left(B,\hat{B} \right)=\sum\limits_{l=1}^{L}\left\|\hat{B}^{(l)}-B^{(l)}\right\|_F^2=\sum\limits_{l=1}^{L}\sum\limits_{i=1}^{N}\sum\limits_{j=1}^{N}\left(\hat{b}_{ij}^{(l)}-b_{ij}^{(l)} \right)^2.$

\subsection{Mixed Graph Embedding Algorithm for Multi-layer Network Community Detection}

We propose a mixed graph embedding-based modularity reconstruction algorithm (MGE-MTR), which incorporates a feature fusion operation into the encoder output. This contrasts with the independent feature approach proposed in the IGE-MTR algorithm described in \autoref{subsection:Identity Graph Tensor Representation Features}.
To be precise, we concatenate the output of the second graph convolutional module $h^{(1)}=GCN_1^2\left(A^1hW_1^1\right)$,$\cdots$,$h^{(L)}=GCN_L^2\left(A^LhW_2^L\right)$ into $N\times(rL)$ dimensional graph representation matrices: $H=concat\left( h^{(1)},h^{(2)},\cdots,h^{(L)}\right)$.

\autoref{optimal low rank reconstruction2} provide the theoretical insurance for converting problem \eqref{single obective maximization} into finding the optimal reconstruction of matrix $M=\alpha_1B^{(1)}+\alpha_2B^{(2)}+\cdots+\alpha_LB^{(L)}$. We try to construct each layer of modularity tensor through a single graph representation matrix $H$.

Similarly, we evaluate the effectiveness of modularity reconstruction through the Frobenius-norm distance between matrix $M$ and the reconstruction matrix $\hat{M}$: $\left\|M-\sigma\left(H\cdot H^T \right) \right\|_F^2$. Intrigued by inequation \eqref{MGRF}, minimization of reconstruction distance $\hat{M}$: $\left\|M-\sigma\left(H\cdot H^T \right) \right\|_F^2$ can be relaxed into the sum of all layers Frobenius-distance $\left\|B^{(m)}-\sigma\left(H\cdot H^T \right) \right\|_F^2,\;\forall m,\;1\le m\le L.$ 

\begin{figure}[H]
	\centering
	\begin{minipage}[b]{0.9\textwidth}
		\includegraphics[width=\textwidth]{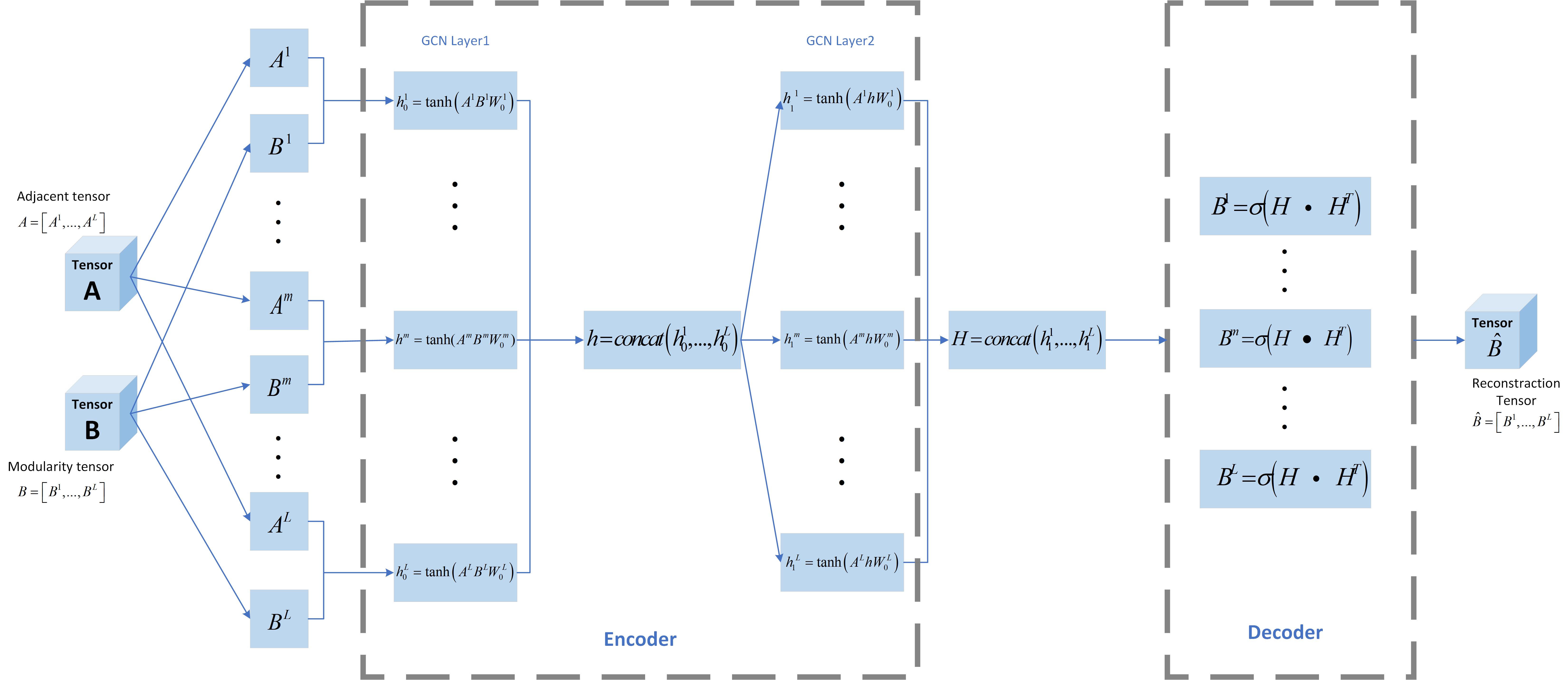}
		\caption{Multi-layer network community detection (MGE-MTR)}
		\label{fig:Mixed Graph Representation Features}
	\end{minipage}
\end{figure}
\autoref{fig:Mixed Graph Representation Features} depicts the framework of independent graph embedding-based multi-layer network community detection. Accordingly, the setting of loss function is $L\left(B,\hat{B} \right)=\sum\limits_{l=1}^{L}\left\|\hat{B}^{(l)}-B^{(l)}\right\|_F^2\\=\sum\limits_{l=1}^{L}\sum\limits_{i=1}^{N}\sum\limits_{j=1}^{N}\left(\hat{b}_{ij}^{(l)}-b_{ij}^{(l)} \right)^2.$

\subsection{Intervention Strategy Based on User Influence}
According to the Matthew effect in social networks \cite{perc2014matthew}, highly influential users play a dominant role in shaping the polarization of group opinions. If their positions are effectively leveraged, they can influence the attitudes of others, thereby achieving the desired intervention outcomes. In this part, we proposed a novel intervention strategy that can be programmatically executed to target communities trapped in information cocoons.
\begin{figure}[H]
	\centering
	\begin{minipage}[b]{0.75\textwidth}
		\includegraphics[width=\textwidth]{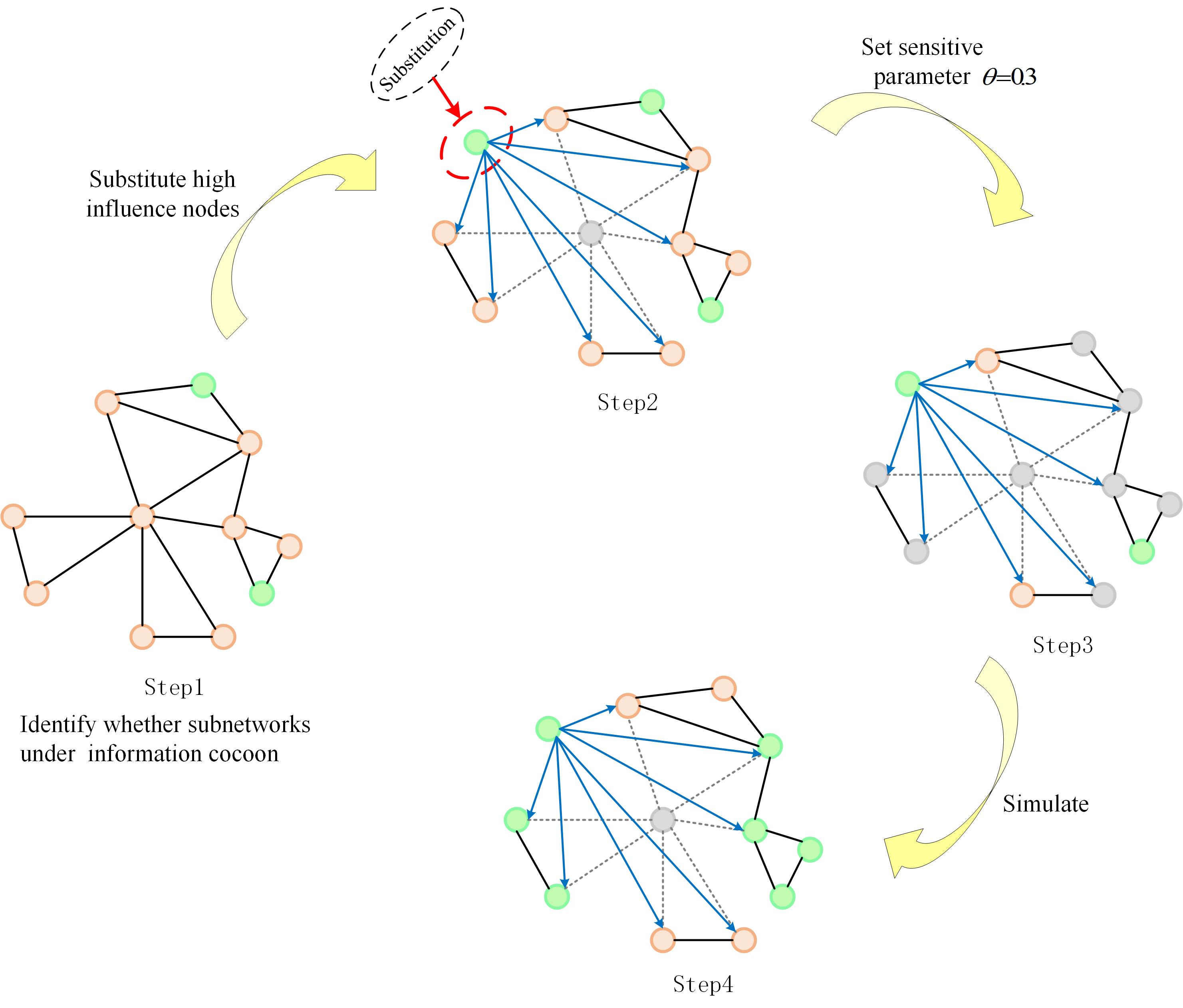}
		\caption{Intervention strategy}
		\label{fig:intervention strategy}
	\end{minipage}
\end{figure}
\autoref{fig:intervention strategy} vividly demonstrates the process of intervention strategy. At first, several highly influential vertices were picked up in a relatively closed sub-network via the influential factors computation method in \autoref{Appendix: Influential Factors}. Try to reduce the exposure of high-impact viewpoints and simultaneously replace those leader attitudes with opposite or neutral viewpoints following the previous recommendation.

The intervention strategy can be simulated programmatically. Suppose that certain users maintain firm attitudes that are resistant to external influence. Consequently, we set a susceptible parameter $\theta$ representing the proportion of users who find it difficult to accept other heterogeneous viewpoints. This group of users always accept other opinions with a small probability. In the simulation process, the users' acceptance of new viewpoints matters, rather than the indiscriminate propagation of viewpoints in the network. User correlations can influence sensitivity, as it is widely acknowledged that users with strong correlations are more likely to exhibit similar sensitivity levels.

\subsection{Simulation System for Double-layer Networks}\label{subsection: Simulation System for Double-layer Network}
This section introduces a simulation system for viewpoint propagation tailored explicitly for a double-layer network scenario. The network comprises two distinct layers: one dedicated to propagating viewpoints and another governing susceptibility status. 

The bottom layer of \autoref{fig:multi-layer network} is constructed by real interaction and response. Moreover, we assume that the viewpoints are propagated in this layer. The nodes in this layer can be classified into three statuses: positive, neutral, and negative. 

Under normal circumstances, two users with similar characteristics are more likely to exhibit comparable acceptance levels. Therefore, we categorize the nodes in the upper layer of \autoref{fig:multi-layer network} into susceptible and insusceptible statuses. These two susceptibility states are dynamically adjusted through the feature similarity network. 

Assume that users in susceptible states are more likely to be influenced by the viewpoints of their neighboring nodes and shift their opinions. In contrast, users with neutral views are more likely to experience changes in susceptibility. 

Motivated by the microscopic Markov chain approach in the research of green behavior propagation among different groups \cite{ziolo2022modeling}, we establish a Markov chain-based state transition simulation system.
\begin{figure}[H]
	\centering
	\begin{minipage}[b]{0.98\textwidth}
		\includegraphics[width=\textwidth]{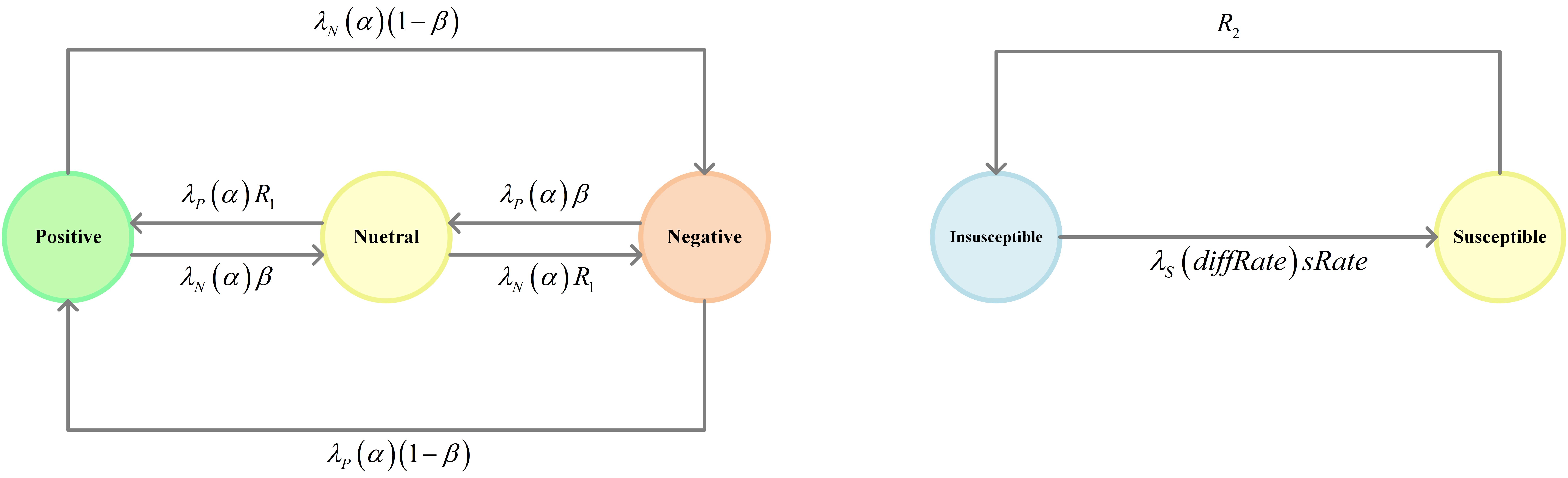}
		\caption{States transition model}
		\label{fig:states transition}
	\end{minipage}
\end{figure}
\autoref{fig:states transition} exhibits how the states of nodes transition in two different layers. The left sub-figure depicts the transition probability among different attitude states of nodes. Parameter $\alpha$ represents the propagation rate, which is the probability that users with different viewpoints spread their views to others. Parameter $\beta$ represents the acceptance rate, which is the probability that users alter their views under the influence of other users. Parameter $R_1$ represents probability of users under neutral status changing their attitudes. We recommend setting the value of $R_1$ to be less than 0.5 to ensure the validity of probability constraints. We define $\lambda$ as propagation adjustment parameters, whose value is affected by propagation rate $\alpha$, inter-layer influential parameters $\gamma_1$ and number of heterogeneous perspectives among adjacent nodes. 
\begin{align*}
	\lambda_{P}^i(\alpha)=1-(1-\gamma_1\alpha)^{N_{positive}}=1-\prod_{j}\left(1-\gamma_1\alpha a_{ij}\mathbbm{1}\left\lbrace attitude[j]==positive \right\rbrace  \right),\\ 
	\lambda_{N}^i(\alpha)=1-(1-\gamma_1\alpha)^{N_{negative}}=1-\prod_{j}\left(1-\gamma_1\alpha a_{ij}\mathbbm{1}\left\lbrace attitude[j]==negative \right\rbrace  \right).
\end{align*}
The nodes with susceptible status in similarity networks are more susceptible to external opinions; meanwhile, the inter-layer interaction influence coefficient is $\gamma_1>1$. When the status of nodes in a similarity network is insusceptible, we choose inter-layer interaction influence coefficient $\gamma_1=1$. In general cases, the range of the inter-layer interaction influence coefficient is $\gamma_1\in\left[1,\frac{1}{\alpha}\right]$.

The right sub-figure is the state transition diagram in the user similarity network. Parameter $R_2$ represents the recovery rate, which is the probability of states transferring from susceptible to insusceptible. Parameter $sRate$ represents the probability of sensitivity changing. Parameter $diffRate$ means transmission rate, which is the probability that the susceptible nodes spread their status. $\lambda_s(diffRate)$ represents the probability of insusceptible nodes changing their states, whose value is influenced by propagation rates, inter-layer interaction coefficient $\gamma_2$ and the number of susceptible nodes in its neighborhood. 
\begin{align*}
	\lambda_{S}(diffRate)=&1-(1-\gamma_2diffRate)^{N_{susceptible}}\\
	=&1-\prod_{j}\left(1-\gamma_2\cdot diffRate\cdot a_{ij}\mathbbm{1}\left\lbrace state[j]==susceptible \right\rbrace  \right).
\end{align*}
When the nodes with neutrality states in the correlation layer, their attitudes tend to polarize more. Thus, their sensitivity would be more likely to alter, at this time, inter-layer interaction influence coefficient $\gamma_2>1$. If the states in the correlation layer are not neutral, we take the value of inter-layer interaction influence coefficient $\gamma_2=1$. In general cases, the range of inter-layer interaction coefficient is $\gamma_2\in\left[1,\frac{1}{diffRate}\right].$
\begin{algorithm}[H] 
	\caption{Framework of Intervention Strategy} 
	\label{alg:Intervention Strategy} 
	\begin{algorithmic}[1] 
		\STATE Set the network status parameters.\\
		Intervention parameters $\eta$, susceptible parameters $\theta$.\\
		The first layer parameters: propagation rate $\alpha$, acceptance rate $\beta$, transition rate $R_1$, inter-layer interaction coefficients $\gamma_1$.\\
		The second layer parameters: propagation rate $diffRate$, sensitivity alteration rate $sRate$, recover rate $R_2$, interlayer interaction coefficients $\gamma_2$.\\
		\STATE Input: Two layer adjacent tensor $A$, Scale of network $N$, Attitude labels.\\
		\STATE Initialize the network states.\\
		$top\_rank\_index=EC[1:\eta N]$,\\
		$l_1\_states^{(0)}=init\_state\left(attitude[top\_rank\_index] \right) $, \\
		$l_2\_states^{(0)}=init\_state(\theta)$.
		
		\STATE Iteration:\\
		While $t<epoch:$\\ 
		$\qquad l_1\_states^{(t+1)}=diffusion1\left(A_1,l_1\_state^{(t)},l_2\_state^{(t)},\alpha,\beta,\gamma_1 \right) $,\\
		$\qquad l_2\_states^{(t+1)}=diffusion2\left(A_2,l_2\_state^{(t)},l_1\_state^{(t)},diffRate,sRate,\gamma_2 \right) $,\\
		$\qquad t=t+1$,\\
		Return $\quad l_1\_states^{(t+1)},l_2\_states^{(t+1)}$.
	\end{algorithmic}
\end{algorithm}

\section{Numerical Experiments}\label{section: Numerical Analysis}

\subsection{Evaluation Index}
\begin{itemize}
	\item Normalized mutual information ($NMI$)\\ 
	To quantify the similarity between actual community affiliations and those identified by algorithms, $NMI$ was introduced for graph community evaluation in 2005 \cite{danon2005comparing}:
	\begin{align*}
		NMI(Y,C)=\frac{2I(Y;C)}{H(Y)+H(C)},
	\end{align*}
	where $Y$ represents the priori class labels of nodes, while $C$ represents the label results processed by the algorithm. $H(\cdot)$ means the cross entropy: $H(X)=-\sum\limits_{i=1}^{|X|}P(i)logP(i)$. $I(Y;C)$ is mutual entropy: $I(Y;C)=H(Y)-H(Y|C)$. In its discrete form, for two different community partitions, $NMI$ can be expressed as follows:
	\begin{align}
		NMI=\frac{-2\sum\limits_{u=1}^{N_A}\sum\limits_{v=1}^{N_B}M_{uv}\log(\frac{nM_{uv}}{M_uM_v})}{\sum\limits_{u=1}^{N_A}M_u\log(\frac{M_u}{n})+\sum\limits_{v=1}^{N_B}M_v\log(\frac{M_v}{n})},
	\end{align}
	where $n$ is the number of graph nodes, $M_{uv}$ is the elements confusion matrix $M$, $N_A$ is the number of communities in partition $A$ and $N_B$ is the number of communities in partition $B$. Besides, $M_u$ is the sum of the $u$-th row of the confusion matrix, and $M_v$ is the sum of the $v$-th row of the confusion matrix. The larger the $NMI$ value, the greater the similarity between the two community structures. If $NMI$ reaches 1, the community partitions are identical.

	\item Modularity degree $Q$ for multi-layer networks \\
	Due to the absence of priori labels for communities in real-world network data analysis, we cannot directly calculate the classification accuracy, thus utilizing modularity degree index $Q$ to evaluate the quality of network partitioning. The value of modularity degree index $Q$ always lies in $[-\frac{1}{2},1]$. If the partition is well-effective, the modularity value goes to $1$. The definition of multi-layer networks with inter-layer coupling is proposed in 2010 \cite{mucha2010community}:
	\begin{align}
		Q=\frac{1}{2M}\sum\limits_{ijsr}\left[\left(A_{ij}^{(s)}-\gamma_s\frac{k_i^{(s)}k_j^{(s)}}{2L^{(s)}} \right)\delta_{sr}+\delta_{ij}\zeta_{jsr} \right]\delta(z_{is},z_{jr}),
	\end{align}
	where $A^{(s)}$ represents the adjacent matrix of $s$-th layer multilayer network and $\zeta_{isr}$ is the index indicator of inter-layer links. $\zeta_{isr}=1$ means the $i$-th node has an edge between $s$-th layer and $r$-th layer. The indicator's value is zero if the inter-layer coupling edge does not exist. $k_i^{(s)}=\sum_{j}A_{ijs}$ represents the total number of edges connected to the $i$-th node in the $s$-th layer. $M$ denotes the total number of layers. $\zeta_{is}=\sum_{r}\zeta_{isr}$ represents total number of links between the $i$-th vertex in $s$-th layer and the $i$-th vertex in the other different layers. $\gamma_s$ is the tuning parameter controlling the expected modularity degree. $\delta(z_{is},z_{jr})$ is the delta indicator that takes the value 1 if $z_{is}=z_{jr}$, and 0 otherwise. If $\delta(z_{is},z_{jr})=1$, the vertex $i$ and vertex $j$ affiliate the same community. Otherwise, the vertex $i$ and vertex $j$ belong to different communities.
	
	Without considering inter-layer coupling connection, \cite{paul2021null} promote Newman-Girvan (NG) modularity degree through layer-wise normalization:
	\begin{align}
		Q_{NM}=\frac{1}{M}\sum\limits_{s}\sum_{i,j}\frac{1}{2L^{(s)}}\left[A_{ij}^{(s)}-\frac{k_i^{(s)}k_j^{(s)}}{2L^{(s)}}\right]\delta(z_i,z_j).
	\end{align}
	It is named multi-normalized
	average (MNavrg) modularity.
	
	Meanwhile, another shared degree modularity degree is proposed in \cite{paul2021null} with the average frequency for estimating links between entity $i$ and entity $j$ in the $s$-th layer, which is given by
	\begin{align}
		Q_{SD}=\frac{1}{M}\sum\limits_{s}\sum_{i,j}\frac{1}{2L^{(s)}}\left[A_{ij}^{(s)}-\frac{L^{(s)}\sum_{s}k_i^{(s)}\sum_{s}k_j^{(s)}}{2L^2}\right]\delta(z_i,z_j).
	\end{align}
	
	\item Similarity index based on KL divergence\\
	All the above evaluation indices are constructed by an adjacent matrix, which merely considers network topology structure. In real-world scenarios, the similarity between node features is often more crucial than the density of topological connections when evaluating the effectiveness of community partitioning. Therefore, we introduce a new evaluation metric based on KL divergence.
	
	KL divergence is a metric that evaluates the similarity between two probability distributions. The discrete form of KL divergence is expressed as
	\begin{align*}
		KL[P(X)||Q(X)]=\mathbb{E}_{X\sim P(x)}\left(log\frac{P(x)}{Q(x)} \right)=\sum\limits_{i=1}^{N}P_ilog\frac{P_i}{Q_i}.
	\end{align*}
	In the community detection task, we aim to ensure that users within the same community are more similar while users in different communities exhibit lower similarity. The similarity evaluation between $i$-th vertex and $j$-th vertex based on KL divergence can be expressed as follows:
	\begin{align*}
		sim_{ij}=KL(H_i||H_j)=\sum\limits_{k=1}^{F}h_k^i\log(\frac{h_k^i+1}{h_k^j+1}),
	\end{align*} 
	where $H_i=\left( h_1^i,\cdots,h_k^i,\cdots,h_F^i\right) ^T$ and $H_j=\left( h_1^j,\cdots,h_k^j,\cdots,h_F^j\right) ^T$ denote the feature vectors of the $i$-th and $j$-th user vertices, respectively. $F$ is the length of the feature vector. By traversing all nodes within the same community $C_k$, the average pairwise similarity among users can be regarded as the similarity index of community $C_k$.
	\begin{align*}
		similarity(C_k)=\frac{1}{N_k}\sum\limits_{i=1}^{N_k}\frac{\sum\limits_{j=1}^{N_k}sim_{ij}}{N_k}=\frac{1}{N_k^2}\sum\limits_{i=1}^{N_k}\sum\limits_{i=1}^{N_k}K\left(H_i||H_j \right),
	\end{align*}
	where $N_k$ is node number in the community $C_k$.  The similarity of the entire networks with K partitions $\{C_1,C_2,\cdots,C_K\}$ is the average of each community similarity indices $similarity(C_k)$:
	\begin{equation}
		\begin{split}
			KL\_similarity\_index(C_1,\cdots,C_K)=&\frac{1}{K}\sum\limits_{k=1}^{K}similarity(C_k)\\
			=&\frac{1}{K}\sum\limits_{k=1}^{K}\frac{1}{N_k^2}\sum\limits_{i=1}^{N_k}\sum\limits_{i=1}^{N_k}K\left(H_i||H_j \right).
		\end{split}
	\end{equation}
	This evaluation index measures the discrepancy between the two distributions of users' features. If features of different users have a significant discrepancy gap, the distance between two samples might be huge, while the value of KL divergence might be large. Thus, the more similarity of users' features within one community, the less value of similarity index $similarity(C_k)$, and the corresponding whole partition evaluation will become less, which means the partition is effective.
	
	\item  Similarity index based on JS divergence\\
	Similar to KL divergence, JS divergence measures the similarity of two probability distributions. Unlike the asymmetry of KL divergence, however, JS divergence has a symmetrical characteristic. 
	\begin{align*}
		JS[P(X)||Q(X)]=\frac{1}{2}KL[P(X)||M(X)]+\frac{1}{2}KL[Q(X)||M(X)],
	\end{align*}
	where $M(X)=\frac{P(X)+Q(X)}{2}$. 
    
    The discrete version of $M(X)$ is $M_{ij}=\frac{H_i+H_j}{2}=\frac{1}{2}\left(h_1^i+h_1^j,\cdots,h_k^i+h_k^j,\cdots,h_F^i+h_F^j \right)^T .$  The similarity index based on JS divergence:
	\begin{equation}
		\begin{split}
			&JS\_similarity\_index(C_1,\cdots,C_k)\\
            =&\frac{1}{2K}\sum\limits_{k=1}^{K}\frac{1}{N_k^2}\sum_{i=1}^{N_k}\sum_{j=1}^{N_k}\left(KL(H_i||M_{ij})+KL(H_j||M_{ij}) \right) \\
			=&\frac{1}{2K}\sum\limits_{k=1}^{K}\frac{1}{N_k^2}\sum_{i=1}^{N_k}\sum_{j=1}^{N_k}\left(KL\left( H_i||\frac{H_i+H_j}{2}\right) +KL\left(H_j||\frac{H_i+H_j}{2}\right) \right),
		\end{split}
	\end{equation}
	where $KL(H_i||M_{ij})=\sum\limits_{k=1}^{F}h_k^i\log(\frac{2(h_k^i+1)}{h_k^i+h_k^j+2}).$

Likewise, if the community partition is effective, the features of users within the same community tend to be similar, resulting in a lower KL divergence $KL(H_i||M_{ij})$. Consequently, the JS similarity index $JS\_similarity\_index(C_1,\cdots,C_k)$ also exhibits a lower value.
\end{itemize}
\subsection{Dataset}
In real-world scenarios, most of the datasets lack priori labels. We evaluate our algorithm regarding prediction accuracy and topological structure in this part. To assess performance, we compare our algorithms with the tucker decomposition with integrated SVD transformation (TWIST) algorithm \cite{jing2021community} using citation datasets (Cora and CiteSeer) and simulated datasets with prior labels. The details of the TWIST algorithm can be found in \autoref{TWIST}.  

Citation datasets consist of authors and their citation relationships. The Cora dataset consists of 2,708 machine learning-related papers categorized into 7 classes. In numerical analysis, we separate this dataset into two parts: gathering reinforcement learning, rule learning, and theory as dataset Cora1 with 724 papers in it and collecting probabilistic method, theory, case-based, and genetic algorithms with 1493 papers in it as the Cora2 dataset. For the Citeer dataset, we select artificial intelligence, machine learning, and agents, three categories with a total number of 1203 as dataset Citeseer1. 

We construct user relationship-similarity two layer networks through citation network dataset. The first layer is constructed via the citation relationship two authors have, and then there exists one edge between the two authors. We consider the word vector in the content file as a feature vector for constructing a second-layer network-similarity network. The cosine value between two feature vectors reflects the node's similarity. The probability of the existence of an edge between two nodes is $\mathbb{P}\left(a_{ij}=1|\bm{z}_i,\bm{z}_j \right)=cos\left\langle\bm{z}_i,\bm{z}_j \right\rangle=\frac{\bm{z}_i\cdot\bm{z}_j}{\left|\bm{z}_i\right|\cdot\left|\bm{z}_j\right|}.$ 

Since the comparison algorithm TWIST has a significant impact on large-sample classification, we add small-size simulation network samples for testing. The simulation datasets are generated by multilayer mixture stochastic block model \cite{jing2021community}. The parameters are set as follows: the layers number is 3, the communities number is 2, the average degree for each single layer is approximately 10, and the number of nodes is 300 and 400.

\subsection{Experimental Analysis}
We apply three algorithms to six datasets for testing. Through NMI and multi-layer networks modularity degree ($Q_{NM}$,$Q_{SD}$), we evaluate the three algorithms from prediction accuracy and network topology structure two aspects. The codes of this section are available. $^1$
\footnotetext[1]{\href{https://github.com/ysw-git123/Multi-layer_network_community_detection_algorithm.git}{\text{https://github.com/ysw-git123/Multi-layer\_network\_community\_detection\_algorithm.git}}}

\renewcommand{\tablename}{Table}
\begin{table}[H]
	\normalsize
	\centering
	\caption{Numerical Experiments Results}
	\label{simulation t=0.2}
	\vspace{3pt}
	\begin{tabular}{|c|c|c|c|c|}
		\hline
		\multirow{2}{*}{Dataset}&\multirow{2}{*}{Evaluation Index}&Algorithm 1&Algorithm 2&Algorithm 3\\ \cline{3-5} &&TWIST&IGE-MTR&MGE-MTR\\ \hline 
		\multirow{3}{*}{Cora1}&$Q_{NM}$&0.1508&0.2603&\textbf{0.3104}\\
		&$Q_{SD}$&0.1706&0.2344&\textbf{0.3062}\\
		&$NMI$&0.0294&0.3237&\textbf{0.3846}\\ \hline
		\multirow{3}{*}{Cora2}&$Q_{NM}$&0.2603&0.2938&\textbf{0.4011}\\
		&$Q_{SD}$&0.2630&0.2905&\textbf{0.3987}\\
		&$NMI$&0.2876&0.3012&\textbf{0.5676}\\ \hline
		\multirow{3}{*}{Citeseer1}&$Q_{NM}$&0.2022&0.1697&\textbf{0.2663}\\
		&$Q_{SD}$&0.2027&0.1693&\textbf{0.2581}\\
		&$NMI$&0.0294&0.0141&\textbf{0.1953}\\ \hline
		\multirow{3}{*}{\makecell{Simulation Dataset\\ (300 vertices)}}&$Q_{NM}$&0.0991&0.1883&\textbf{0.1919}\\
		&$Q_{SD}$&0.4087&0.4715&\textbf{0.4741}\\
		&$NMI$&0.3130&0.9288&\textbf{1.0000}\\ \hline
		\multirow{3}{*}{\makecell{Simulation Dataset\\ (400 vertices)}}&$Q_{NM}$&0.1043&0.1089&\textbf{0.1952}\\
		&$Q_{SD}$&0.4186&0.3868&\textbf{0.4732}\\
		&$NMI$&0.2632&0.0058&\textbf{1.0000}\\ \hline
		\multirow{3}{*}{\makecell{Simulation Dataset\\ (500 vertices)}}&$Q_{NM}$&\textbf{0.1962}&0.1941&\textbf{0.1962}\\
		&$Q_{SD}$&\textbf{0.4757}&0.4739&\textbf{0.4757}\\
		&$NMI$&\textbf{1.0000}&0.9530&\textbf{1.0000}\\ \hline
	\end{tabular}\label{table: numerical experiments}
\end{table}
\autoref{table: numerical experiments} illustrates that our novel mixture graph embedding-based multi-layer community detection algorithm (MGE-MTR) significantly outperforms other algorithms in community detection tasks. The prediction accuracy in labels and network topological partition effect have been promoted more than previous ones. Meanwhile, our novel MGE-MTR multi-layer community detection algorithm overcomes the drawback of low classified accuracy in small scale networks.

\section{Real Data Analysis}\label{section: Real Data Analysis}
In the social media topic dissemination process, users are prone to find the search points of view that could support their opinions. Under the influence of recommendation systems, new users can find the idea they agree with and join them swiftly. The topic debate group would gradually evolve into two conditions: the one is viewpoints gradually evolving into two opposite versions, and users holding these opinions steadily separate into different camps, forming multiple information cocoons. Another situation is that multiple perspectives existed initially, but with the topic becoming hot, one category viewpoint places the dominant and submerges other viewpoints. Users holding distinctive opinions are gradually inclined to reach a consensus. In some cases, such consensus can be positive, or even extremely one-sided, ultimately resulting in the formation of an entire information cocoon.

In topic comments networks, users' emotional tendencies have specific associations with community structures. However, not all users in one relatively closed social network community necessarily hold the same attitude, which means not all sub-networks have experienced the phenomenon of information cocoons. Sometimes, different opinions exist in one sub-network, which cannot be ascertained as being trapped in information cocoons. Moreover, there is no need to exert intervention measures on those sub-networks. Our task is to explore the sub-network in which the information cocoons might occur. Users with frequent interactions and connections might be divided into the same community, and individuals within the same community are prone to being trapped in information cocoons. In this section, we apply our novel community detection algorithms to explore the potential information cocoons in social networks. The code of real data analysis and Algorithm \ref{alg:Intervention Strategy} is available.$^2$
\footnotetext[2]{\href{https://github.com/ysw-git123/Multi-layer_network_community_detection_algorithm.git}{\text{https://github.com/ysw-git123/Multi-layer\_network\_community\_detection\_algorithm.git}}}

\subsection{Data Pre-Processing}\label{subsection: data pre-processing}
Select the fierce debate topic on social media platforms to investigate information cocoons. In our research, we selected the comments about the movie "Moon Man" and collected some of the topics discussed about this movie from the social platform Weibo. The keywords from the website include username, user reply recipient, comment posted location, user level, and user comment content. 
In order to maintain a single channel of information acquisition and minimize the influence of divergent perspectives from other platforms, we focus on a short time window of user comments on the Weibo platform. It is assumed that, within this period, users obtain information exclusively from Weibo. The selected time frame is from 16:00 to 21:00 on April 10th, 2022.

During dataset preprocessing, we filtered out comments containing fewer than eight words, those consisting solely of emojis or images, and removed samples from low-activity users who participated in the discussion. After preprocessing, a total of 775 comments were retained for analysis.

Use the pre-trained model in the NLTK package to classify sentiment and attitude tendencies in comments , followed by manual adjustment of the outputs for improved accuracy. In the final filtered dataset, 417 comments hold negative attitudes toward the movie, 232 comments maintain supportive attitudes toward the movie, and 128 comments have neutral attitudes about the movie's quality, which we classify as neutral.


\begin{figure}[H]
	\centering
	\begin{minipage}[b]{0.45\textwidth}
		\includegraphics[width=\textwidth]{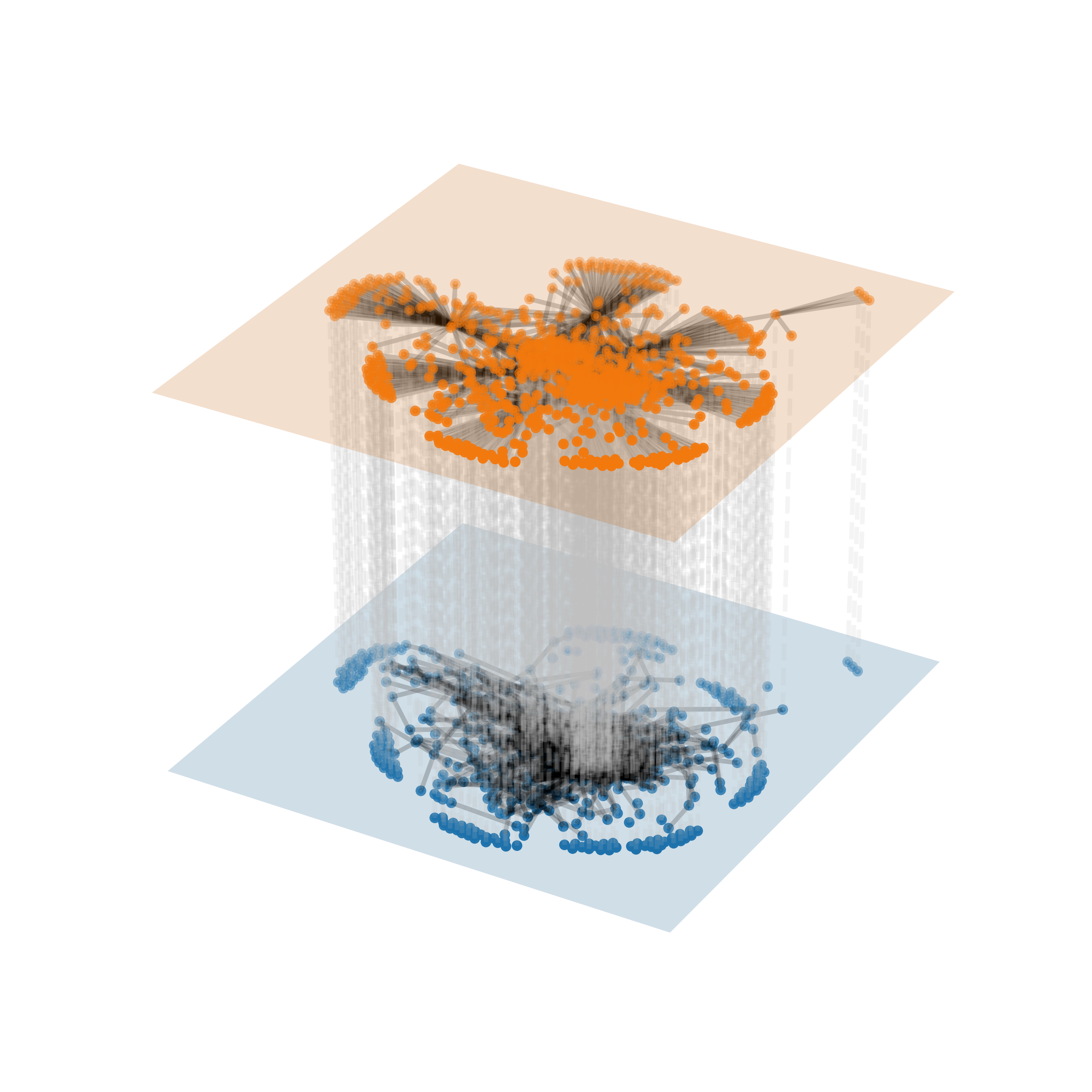}
		\caption{Multi-layer network}
		\label{fig:multi-layer network}
	\end{minipage}
\end{figure}
In \autoref{fig:multi-layer network}, the bottom layer is constructed by reply connection, and the top layer is generated by user similarity. The nodes in the network represent each user, the inter-layer dashed lines mean node alignment, and real lines within one layer mean real connections. The network constructed by the relationship of comments and responses appears to have obvious cluster distribution, meaning that most interactions occur within clusters with less interaction between clusters. Meanwhile, there is a small cluster of vertices with some connections with different clusters and a small number of nodes disconnected from the leading network. Due to the similarity-based recommendation system's impact and other users' expressions, the similarity probability generates the second layer. The similarity is computed using the cosine similarity of the weighted word frequency feature vectors. If the similarity value is high, one edge between two vertices is more likely to exist. A two-layer network reflects the connection of users in different dimensions, which is suitable for exploring information in sparse networks in a single layer.

\subsection{Partition Communities}\label{subsection: Partition Communities}
\begin{figure}[H]
	\centering
	\begin{minipage}[b]{0.55\textwidth}
		\includegraphics[width=\textwidth]{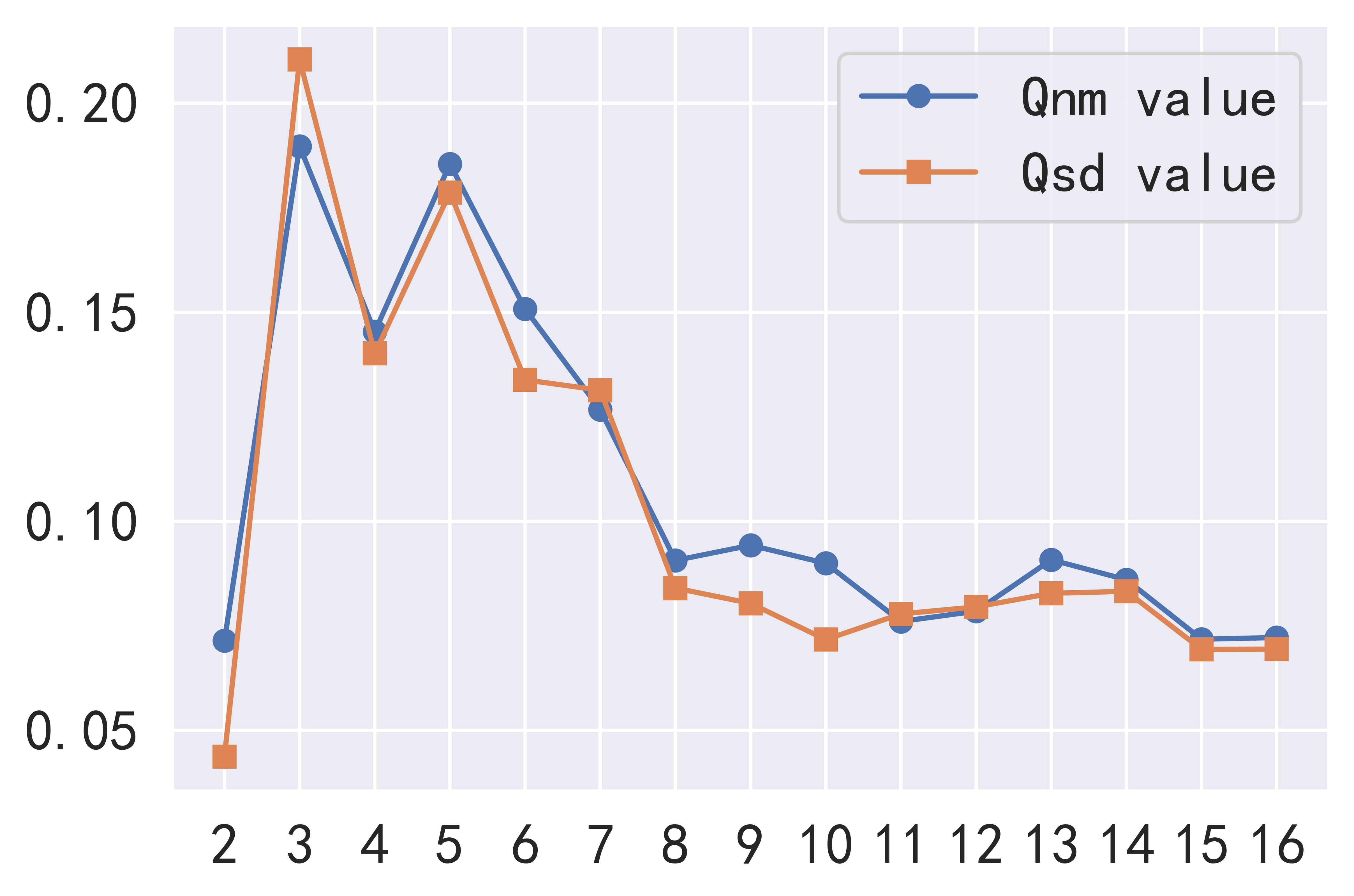}
		\caption{Q value variation with community count.}
		\label{fig:Q value varying with number of communities}
	\end{minipage}
\end{figure}

Exploration of information cocoons aims to partition relatively independent and closed sub-networks without dense connections. Real-world problems always lack priori labels, so we need to determine the optimal number of communities. Here, we leverage our proposed MGE-MTR algorithm and try to find the optimal number of communities corresponding with modularity Q value. Traversing the numbers 2 to 16, \autoref{fig:Q value varying with number of communities} exhibits modularity Q values vary with the number of communities.

\begin{figure}[H]
	\centering
	\begin{minipage}[b]{0.55\textwidth}
		\includegraphics[width=\textwidth]{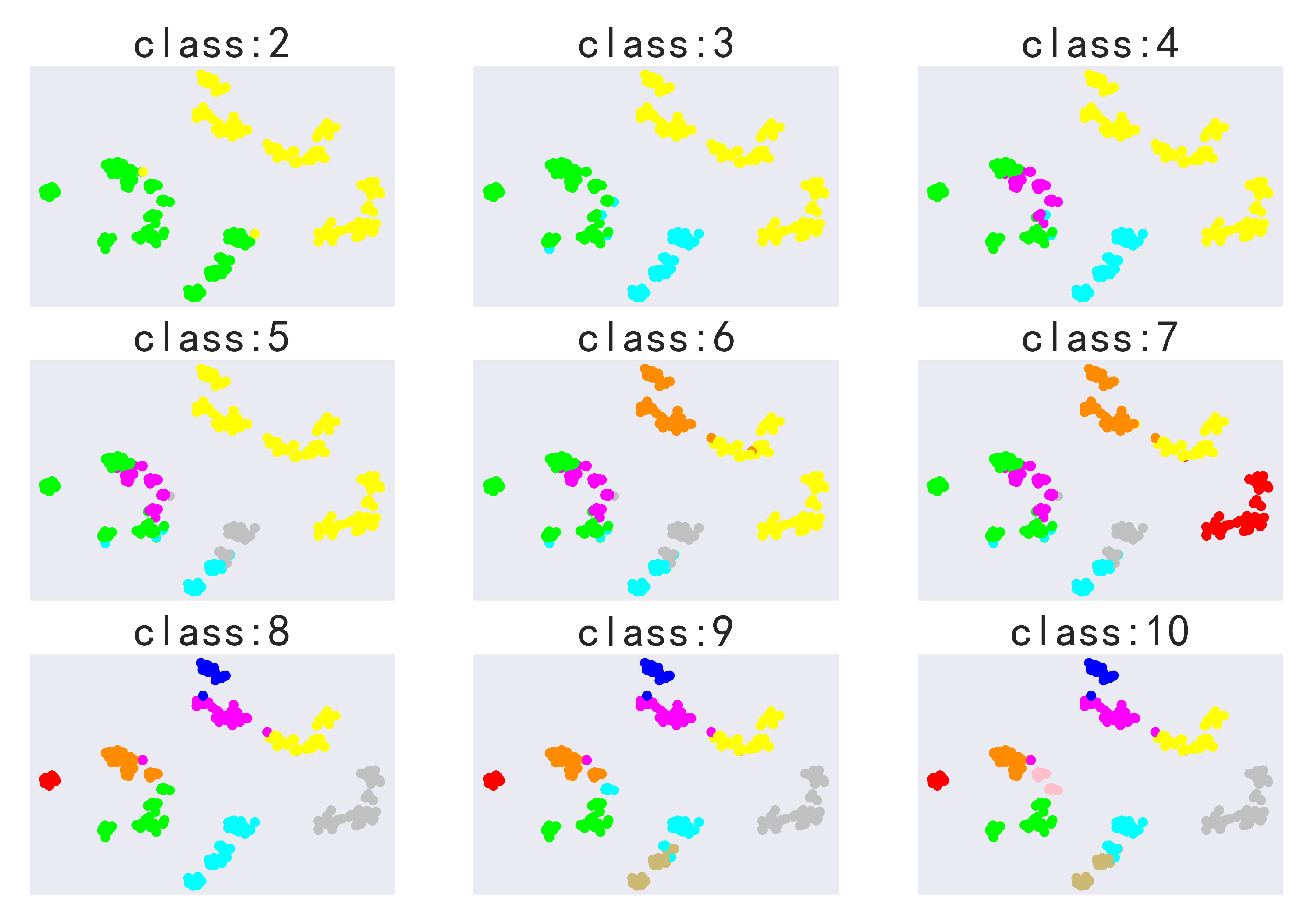}
		\caption{t-SNE dimensionality reduction.}
		\label{fig:t-SNE}
	\end{minipage}
\end{figure}

t-SNE is a widely-used method that could avoid the trend of feature vectors centralizing and maintaining the distance of high-dimension features. Here, we use the t-SNE method to project the high dimensional output vectors into two-dimensional space, and visualization results are exhibited in \autoref{fig:t-SNE} with the communities number ranging from 2 to 10. From \autoref{fig:Q value varying with number of communities}, the optimal partitioning of the topological structure is achieved when the number of communities is set to three. Additionally, the cluster feature map in \autoref{fig:t-SNE} vividly depicts projected features with class labels. When setting three as the community number, features in such partition remain a certain distance without apparently overlapping.
\begin{figure}[H]
	\centering
	\begin{minipage}[b]{0.45\textwidth}
		\includegraphics[width=\textwidth]{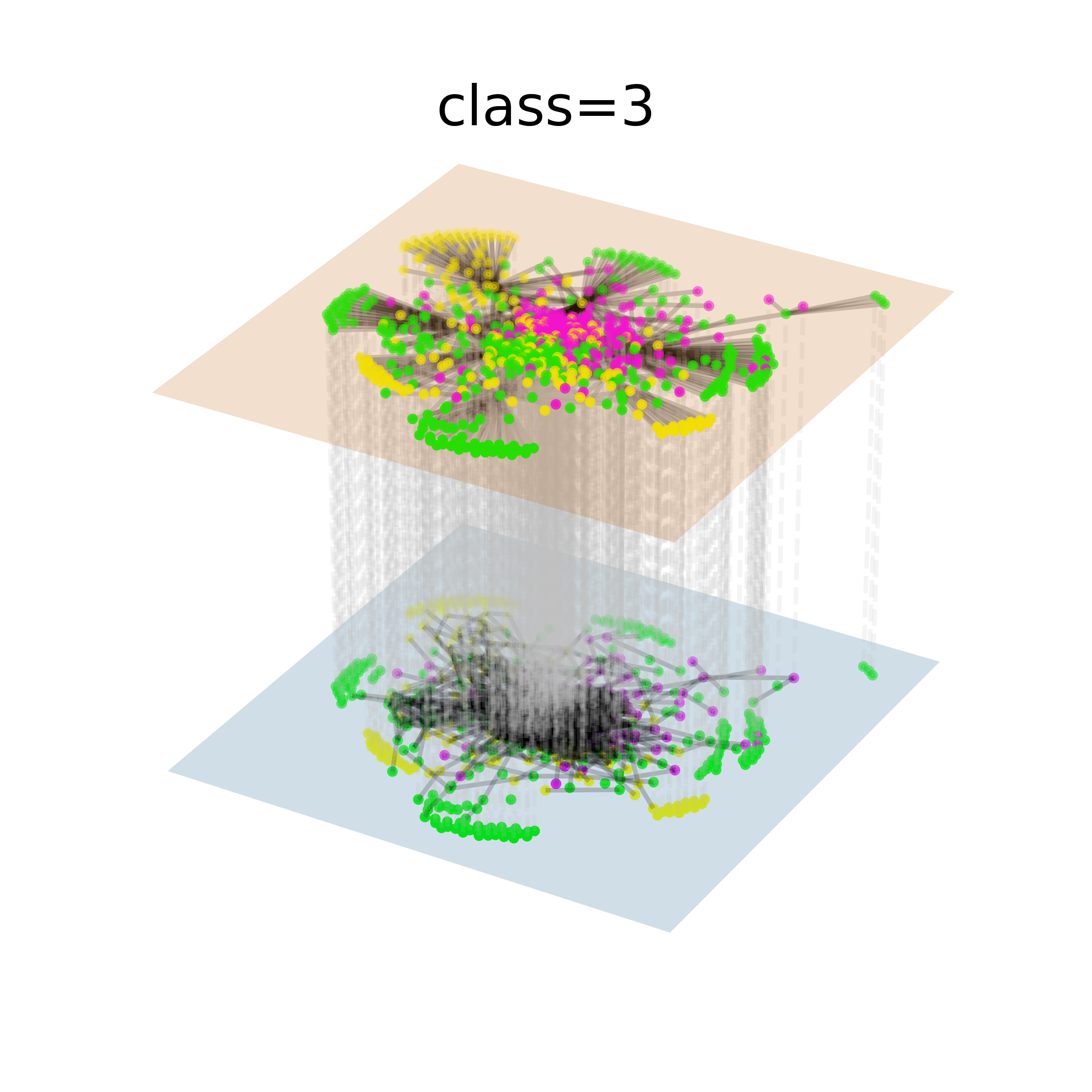}
		\caption{Community partition}
		\label{fig:networks with community partition}
	\end{minipage}
\end{figure}

\autoref{fig:networks with community partition} demonstrates the distribution of three communities, from which we know users belonging to the same communities have more association than those from different communities in terms of similarity or interaction.
The upper layer network reveals the latent association between low-activity-reliant nodes and less interactive user groups within the bottom layer. 

To verify the effectiveness of our double-layer network modeling approach, we compare the similarity-based double-layer network with knn-based double-layer network and single layer network in terms of the topology structure and partition similarity.  \autoref{table: Comparison of networks modeling method} illustrates that our similarity-based network modeling approach is the optimal choice.

\renewcommand{\tablename}{Table}
\begin{table}[H]
	\normalsize
	\centering
	\caption{Comparison of Networks Modeling Method}
	\label{table: Comparison of networks modeling method}
	\vspace{3pt}
	\begin{tabular}{|c|c|c|c|c|c|}
		\hline
		\multirow{2}{*}{\makecell{ Evaluation Index\\Network}}&\multirow{2}{*}{Algorithm}&\multirow{2}{*}{$Q_{NM}$}&\multirow{2}{*}{$Q_{SD}$}&\multirow{2}{*}{$KL_{similarity}$}&\multirow{2}{*}{$JS_{similarity}$}\\ \cline{1-1} & &&&&\\ \hline
		\multirow{3}{*}{\makecell{Similaity-based\\ double-layer\\ network}}&IGE-MTR&0.1623 &0.1728 &\textbf{0.0775}&\textbf{0.0758}\\ \cline{2-6}
		&MGE-MTR&\textbf{0.2586}&\textbf{0.2383}&0.0955&0.0955\\ \cline{2-6}
		&TWIST&0.1572&0.1834&0.1184&0.1064\\ \hline
		\multirow{3}{*}{\makecell{knn-based \\double-layer\\ network (k=6)}}&IGE-MTR&0.1107&0.1234&0.1181&0.1058\\\cline{2-6}
		&MGE-MTR&0.1902&0.2217&0.1192&0.1071\\ \cline{2-6}
		&TWIST&0.0789&0.0551&0.1037&0.0949\\ \hline
		Single layer network&MMR&0.1650&0.1625&0.1121&0.1054\\ \hline
	\end{tabular}
\end{table}

Next, we need to determine which community is trapped in an information cocoon and analyze the three communities that have been divided individually. We compute the percentage of attitude labels in those three communities. In the first community, the negative attitude occupies a dominant position, taking up 77.63 percent of all users in this community. People with those three attitudes constitute one-third of the total in the second community. In the final community, the positive attitude comments comprise approximately two-thirds of all comments. Consequently, the first and last communities are likely to be trapped in an information cocoon. 

To confirm this, take the first community, for example; we select the top 15 highly influential users. The influence factors are calculated by the eigenvector centrality-based method \cite{sola2013eigenvector}, which assumes that the importance of a certain vertex in a complex network relies on itself and its neighbor vertices. If a vertex is connected with a highly influential node, this vertex might obtain high importance. Assume the transferring matrix is $M=D^{-1}A$, where $A=\left(a_{ij} \right)_{N\times N} $ is adjacent matrix of limited graph, $D=diag(d_1,d_2,\cdots,d_N)$ is the degree matrix of adjacent matrix and $d_i=\sum\limits_{j=1}^{N}a_{ij}$ is degree of the $i$-th degree. The eigenvector centrality is composed of the eigenvector centrality of its neighbors and its influence on its around neighbors, which can be formulated as:
\begin{align*}
	EC'(i)=\lambda\sum\limits_{j=1}^{N}m_{ij}EC(j)+(1-\lambda)EC(i),
\end{align*}
whose vector iteration version is:
\begin{align*}
	EC^{(\alpha+1)}=\lambda M\cdot EC^{(\alpha)}+(1-\lambda)EC^{(\alpha)}.
\end{align*}
The explicit influential factors calculation process will be demonstrate in \autoref{Appendix: Influential Factors}.

\autoref{fig:influential factors} depicts the first communities' top 15 highly influential users. As is shown in \autoref{fig:influential factors}, the main attitude in those highly influential users is negative, accounting for over 93 percent.
\begin{figure}[H]
	\centering
	\begin{minipage}[b]{0.58\textwidth}
		\includegraphics[width=\textwidth]{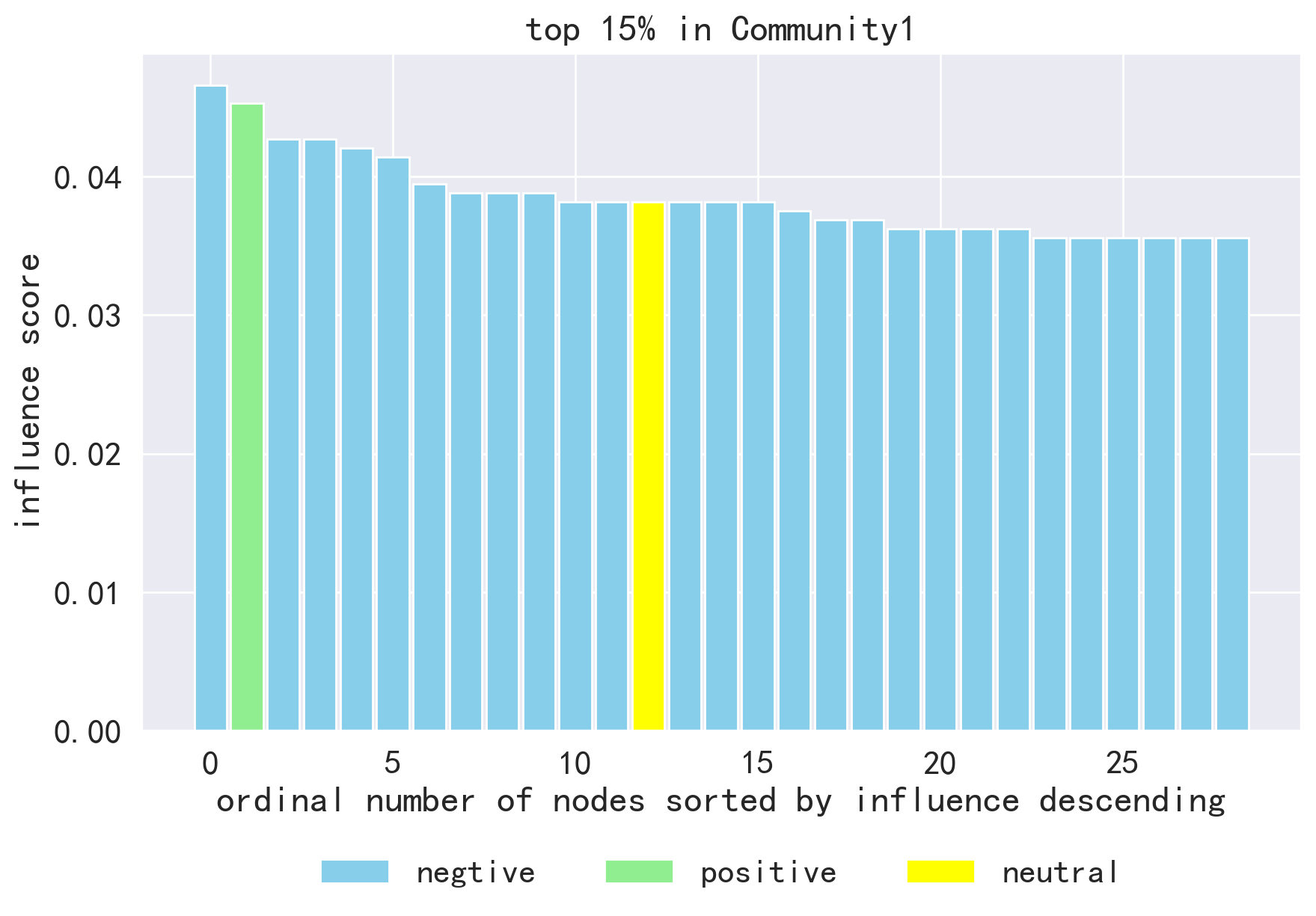}
		\caption{Influential scores of high influential vertex}
		\label{fig:influential factors}
	\end{minipage}
\end{figure}

\subsection{Simulation of Viewpoint Dissemination}
Based on the two-layer network established in \autoref{subsection: data pre-processing}, this part tests the viewpoint propagation simulation system proposed in \autoref{subsection: Simulation System for Double-layer Network}. The multi-layer networks comprise viewpoints propagation layer and susceptibility effect layer.  

For simulation, we set our susceptible parameter $\theta=0.1$.
Firstly, we will test the stability of our simulation system. From 5 percent to 25 percent, we select five groups of intervention ratio with a 5 percent interval. The aim is to examine whether the ratio of each state in two layers will achieve stability after some time in status propagation. The parameters of the simulation system are set as follows: 
\renewcommand{\tablename}{Table}
\begin{table}[H]
	\normalsize
	\centering
	\caption{Numerical Experiments Parameter Setting}
	\label{Numerical Experiments Parameter Setting}
	\vspace{3pt}
	\begin{tabular}{|c|c|c|c|c|c|}
		\hline
		\multicolumn{3}{|c|}{User Relationship layer}&\multicolumn{3}{c|}{User Similarity layer}\\ \cline{1-3}\cline{4-6}
		Parameter&Name&value&Parameter&Name&value\\ \hline
		$\alpha$&propagation rate&0.3&$diffRate$&propagation rate&0.3\\ \hline
		$\beta$&acceptance rate&0.2&$sRate$&susceptible rate&0.2\\ \hline
		$R_1$&transition rate&0.3&$R_2$&recover rate&0.2\\ \hline
		$\gamma_1$&inter-layer coefficients 1&1.5&$\gamma_2$&inter-layer coefficients 2&1.5\\ \hline
	\end{tabular}
\end{table}
In the numerical study, we iterate the diffusion process 50 times and examine whether the ratios of nodes with insusceptible states in similarity networks come to stability.

\begin{figure}[H]
	\centering
	\begin{minipage}[b]{0.55\textwidth}
		\includegraphics[width=\textwidth]{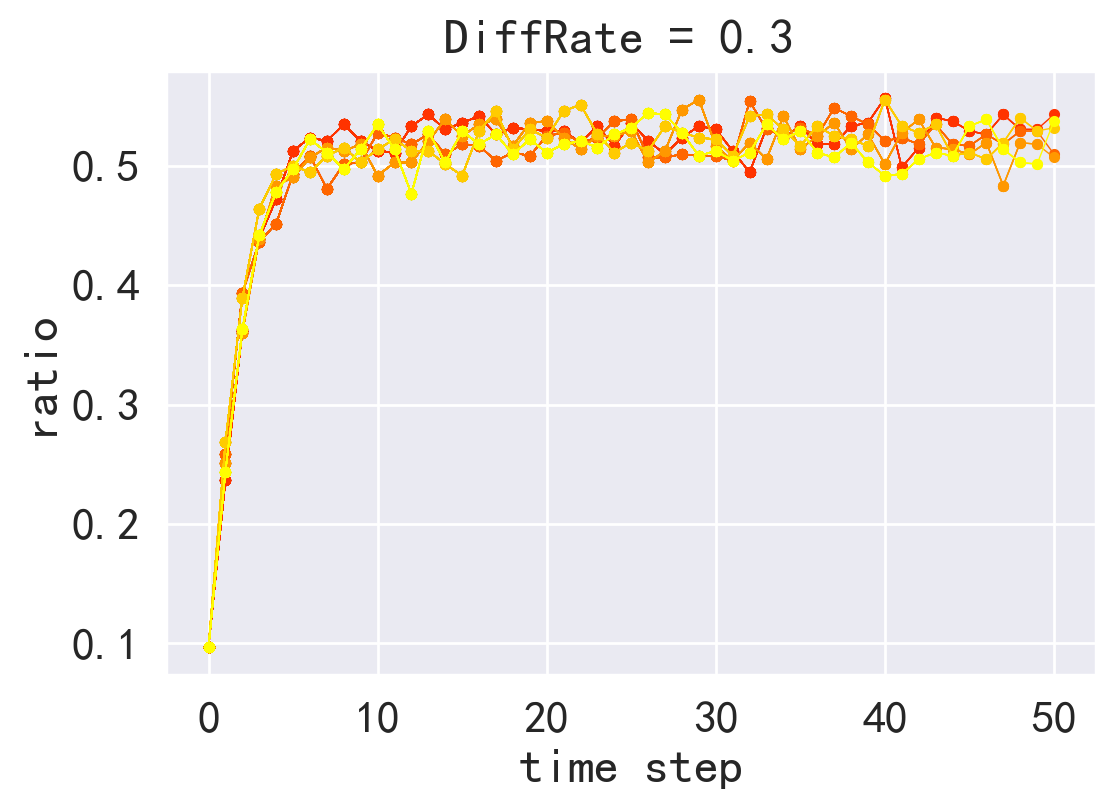}
		\caption{Simulation iteration process}
		\label{fig:iteration process}
	\end{minipage}
\end{figure}
The horizontal axis of  \autoref{fig:iteration process} substitutes iteration steps for attitudes propagation and random status change. In contrast, the vertical axis indicates the share of insusceptible state nodes in a similarity network. With iterations, it is evident that the proportion of insusceptible nodes remains substantially stable at 50 percent with slight fluctuation. We can confidently say that the propagation comes to global relative stability after maximum iterations. At that time, the node's status can be reckoned as the final distribution of information dissemination.

\begin{figure}[H]
	\centering
	\begin{minipage}[b]{0.58\textwidth}
		\includegraphics[width=\textwidth]{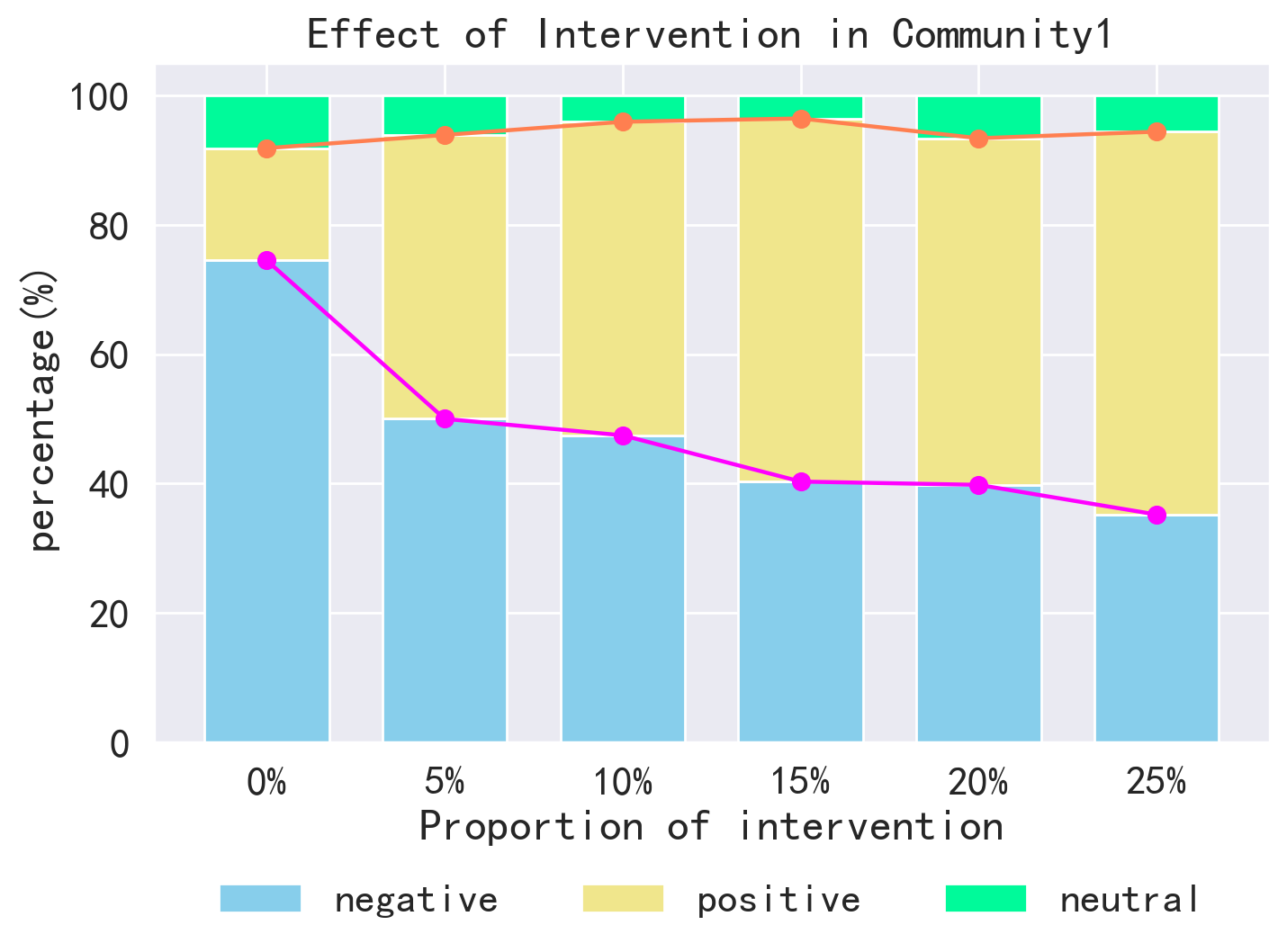}
		\caption{Intervention effect in the first community}
		\label{fig:intervention effect in community1}
	\end{minipage}
	
\end{figure}

\begin{figure}[H]
	\centering
	\begin{minipage}[b]{0.58\textwidth}
		\includegraphics[width=\textwidth]{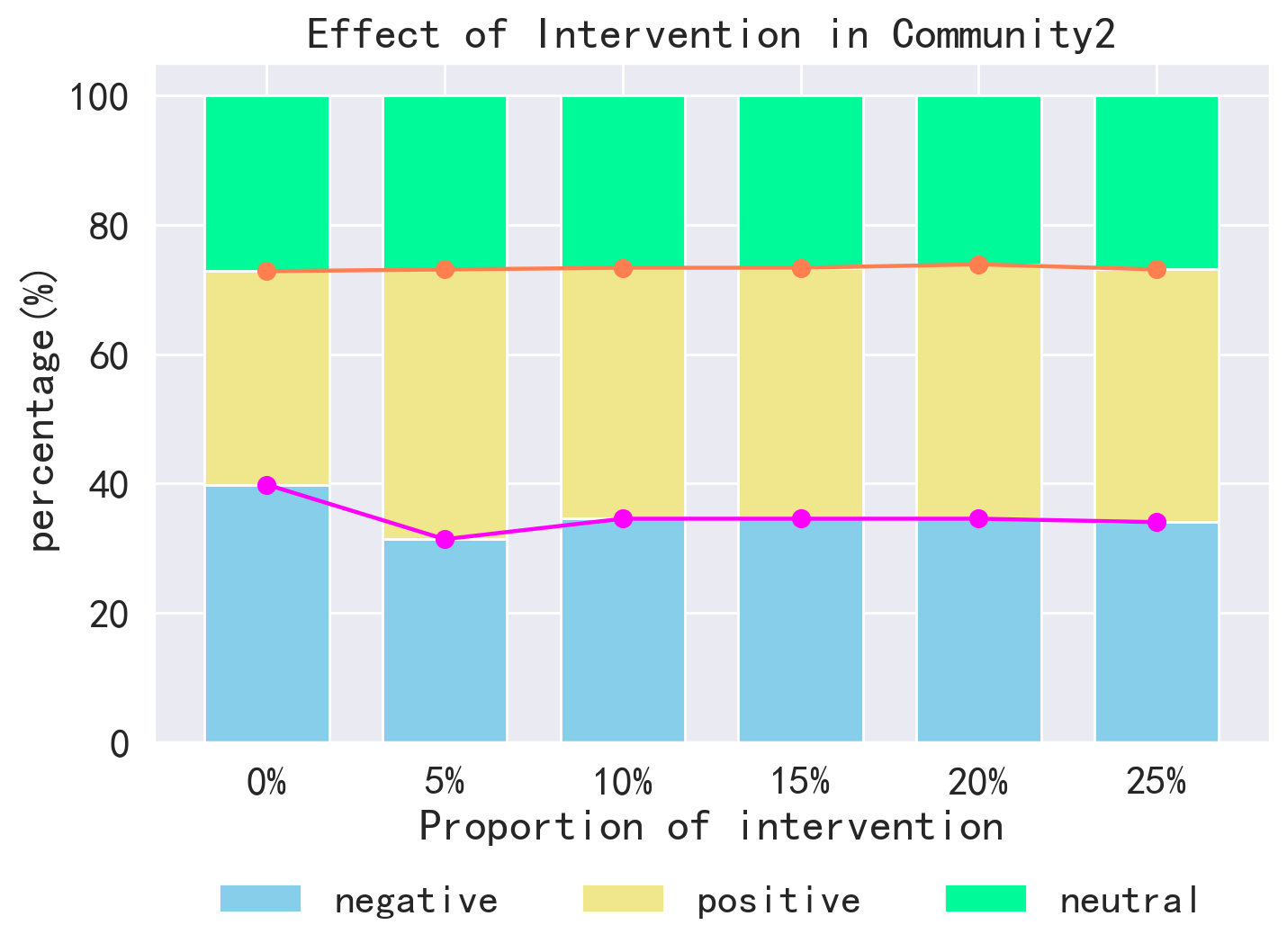 }
		\caption{Intervention effect in the second community}
		\label{fig:intervention effect in community2}
	\end{minipage}
\end{figure}
We try to exert intervention on the first community mentioned in \autoref{subsection: Partition Communities} and simulate the distribution of viewpoints after 50 times iterations under different intervention proportions. The result exhibits in \autoref{fig:intervention effect in community1}.

When the propagation stabilizes, the distribution of attitudes in the first community experiences a significant change. With the enhancement of the intervention level, the rate of positive attitudes shows an upward trend, while the share of negative attitudes decreases significantly. This means that our proposed intervention method can well relieve the polarization of viewpoints.

We also want to check the effect of intervention in other communities without intervention measures. \autoref{fig:intervention effect in community2} exhibits the fluctuation of attitudes label distribution in a non-intervention community. We can conclude that there are no apparent changes in the second community when we only exert intervention measures on the first community. It also illustrates that there is little influence between communities while the standpoints propagate only within the community, which means the community partitions can well explore information cocoons. Our proposed approach is effective in alleviating the emergence of information cocoons.

\section{Discussion} \label{section: Discusion}

\subsection{Sensitivity Analysis for Simulation System}
In the last section, we propose a novel propagation simulation system for two-layer networks. We set two initialization parameters and eight simulation system parameters in that simulation model. In this section, we would like to perform a sensitivity analysis of some parameters to compare the final distribution when the parameters alter. 

We first analyze the sensitivity of two initialization parameters, including sensitivity parameter $\theta$ and intervention ratio $\eta$. Setting five groups of susceptible parameters ranging from 0.1 to 0.5, we traverse the six groups of intervention ratio from 0 to 25 percent and compare the changes in the ratio of users holding negative attitudes.

\begin{figure}[H]
	\centering
	\begin{minipage}[b]{0.59\textwidth}
		\includegraphics[width=\textwidth]{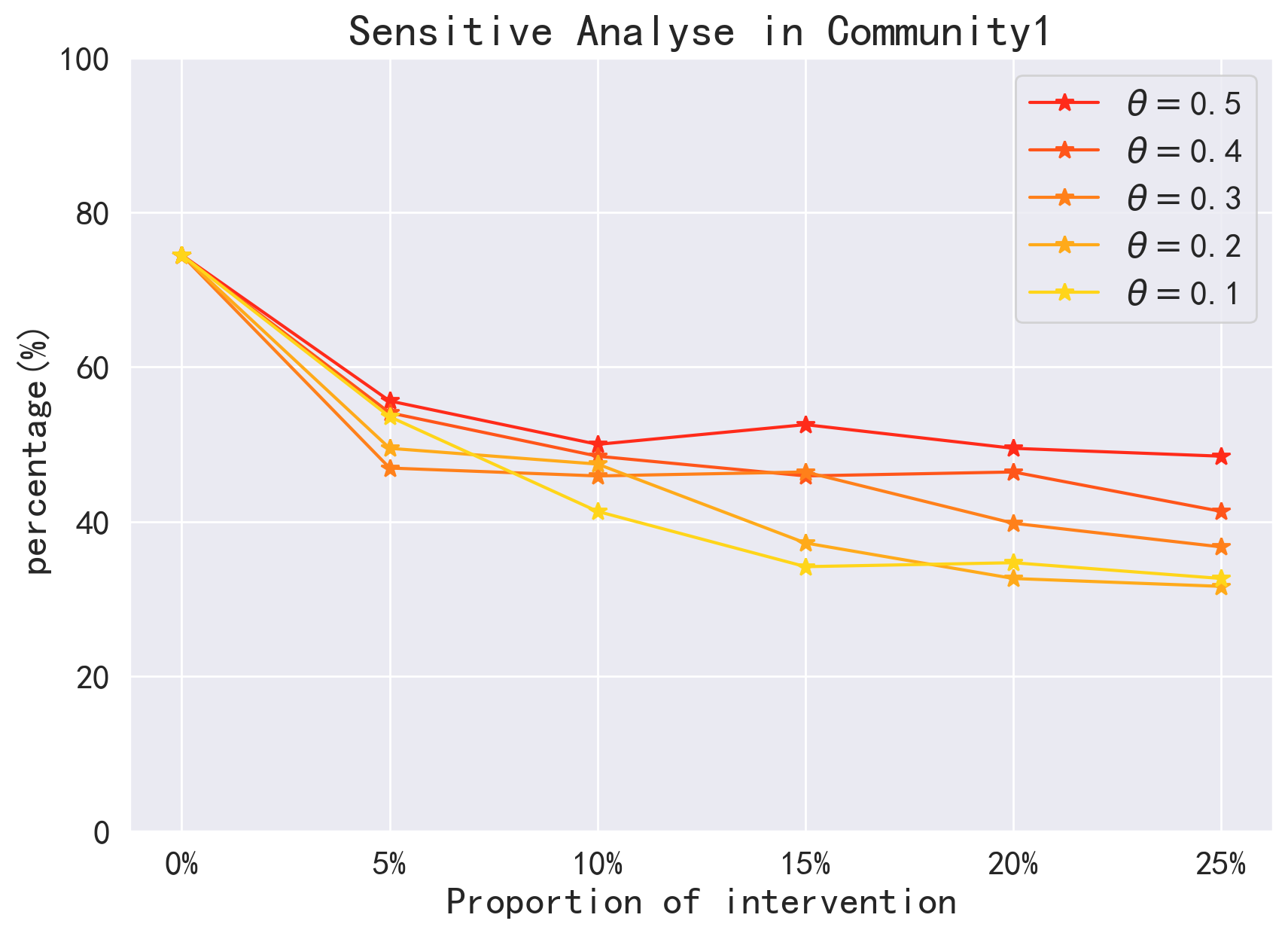}
		\caption{Sensitivity analysis in the first community}
		\label{fig:sensitive analysis  in community1}
	\end{minipage}
\end{figure} 
\begin{figure}[H]
	\centering
	\begin{minipage}[b]{0.59\textwidth}
		\includegraphics[width=\textwidth]{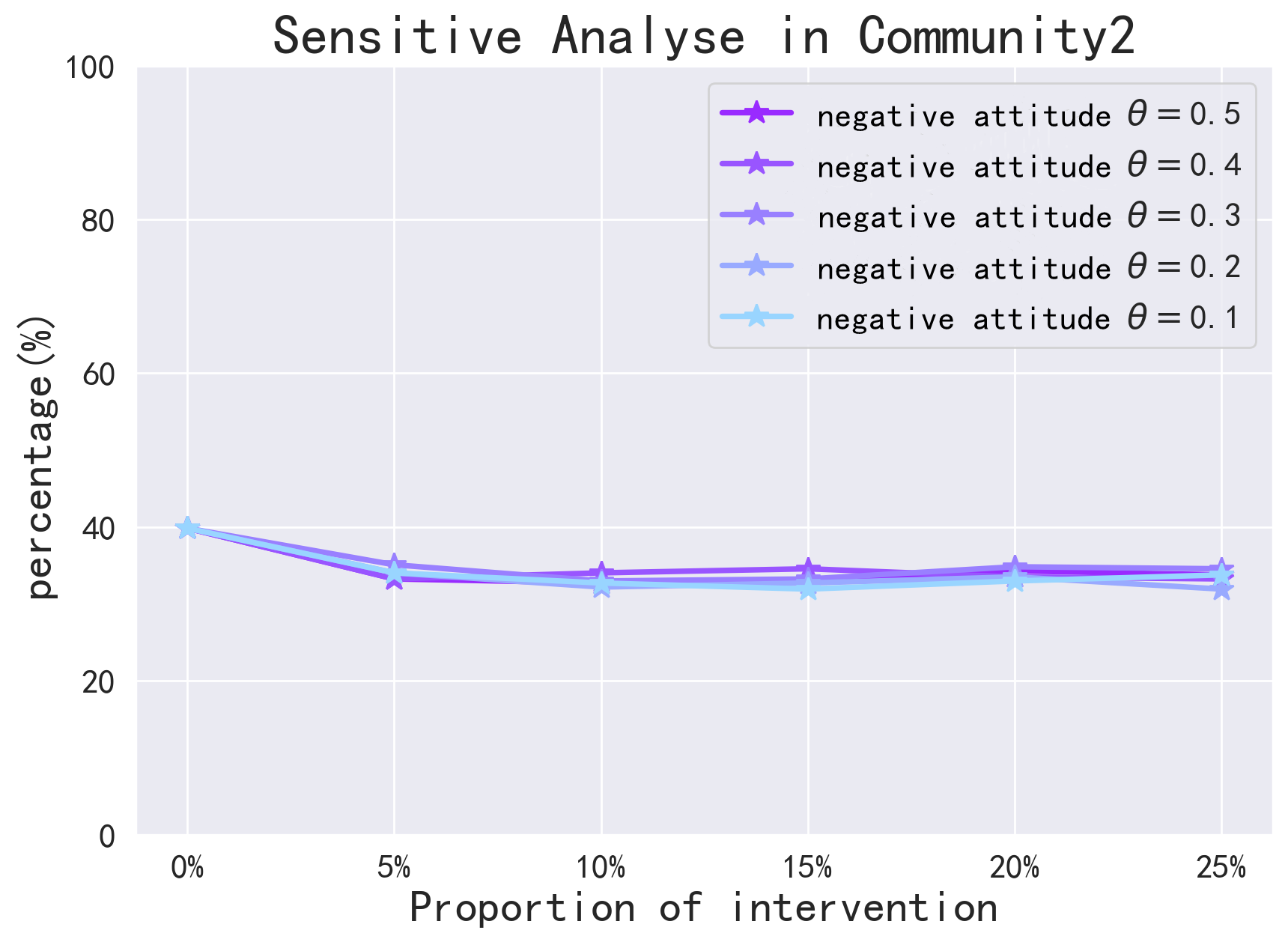}
		\caption{Sensitivity analysis in the second community}
		\label{fig:sensitive analysis in community2}
	\end{minipage}
\end{figure} 
As we can see from \autoref{fig:sensitive analysis in community1}, under the different susceptible ratio $\theta$, the intervention effects vary to some extent in the communities where the intervention measures are mainly implemented, indicating the proposed simulation system  has a certain sensitivity to parameter $\theta$. As mentioned above, the lower the susceptible parameter $\theta$ is, the fewer people insist on their standpoints. The remaining part of people is more receptive to others' viewpoints, under which conditions they are more likely to change their standpoints. Therefore, the intervention effects will be sensitive to intervention intensity, especially as the susceptibility parameter decreases.

As shown in \autoref{fig:sensitive analysis in community2}, the ratio of negative attitudes in the second community is slightly different, with the susceptible parameters $\theta$ changing when laying intervention measures in the first community. As the intervention ratio increases, the proportion of negative attitudes only appears moderate decline and remain unchanged overall.

Next, we analyze the sensitivity of simulation system parameters to examine the influence on propagation system stability with various parameters.

\begin{figure}[H]
	\centering
	\begin{minipage}[b]{0.58\textwidth}
		\includegraphics[width=\textwidth]{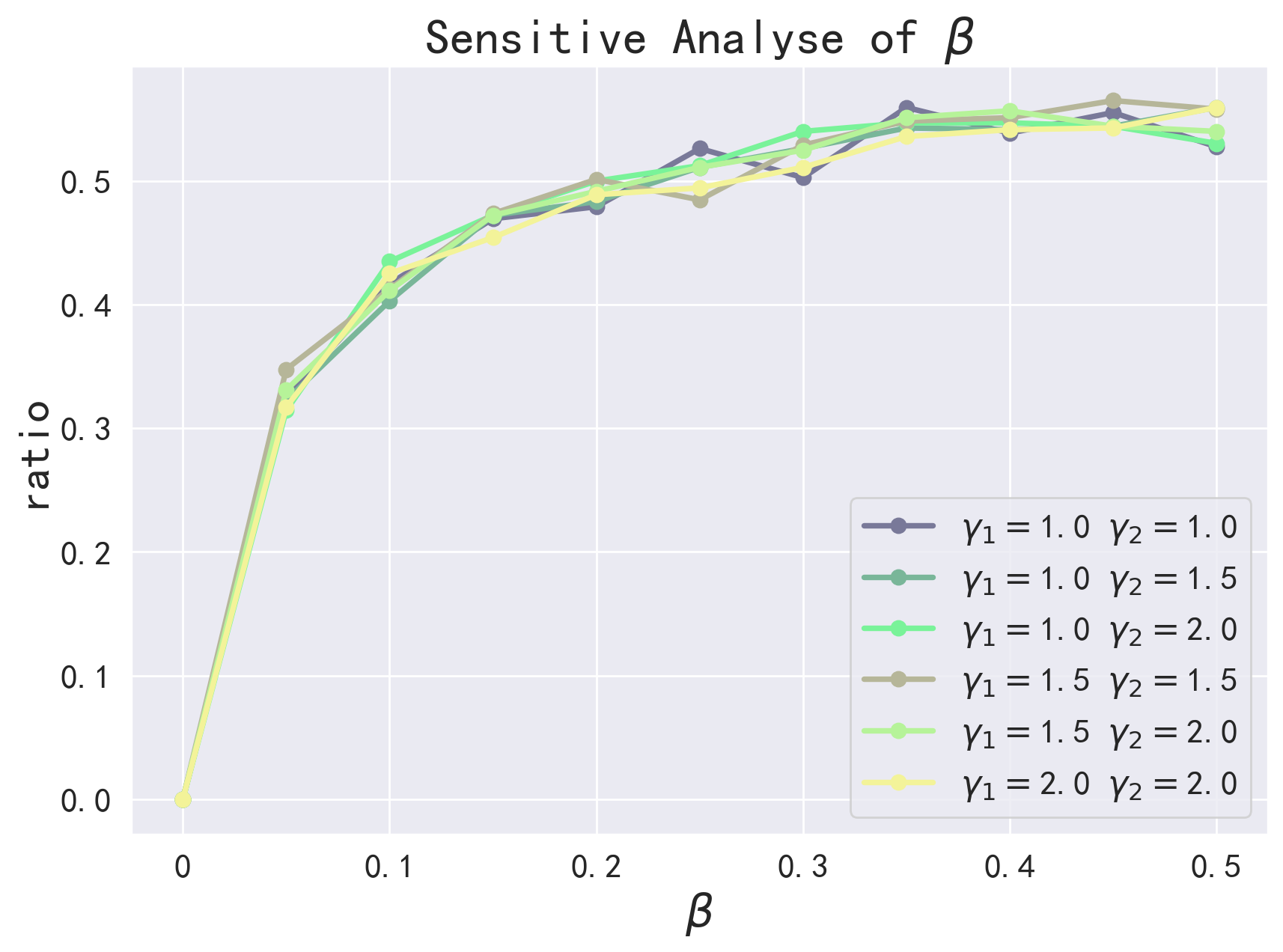 }
		\caption{Sensitivity analysis for $\gamma_1$, $\gamma_2$ and $\beta$}
		\label{fig:sensitive analysis for inter-layer coefficients and acceptance rate}
	\end{minipage}
\end{figure} 

\begin{figure}[H]
	\centering
	\begin{minipage}[b]{0.58\textwidth}
		\includegraphics[width=\textwidth]{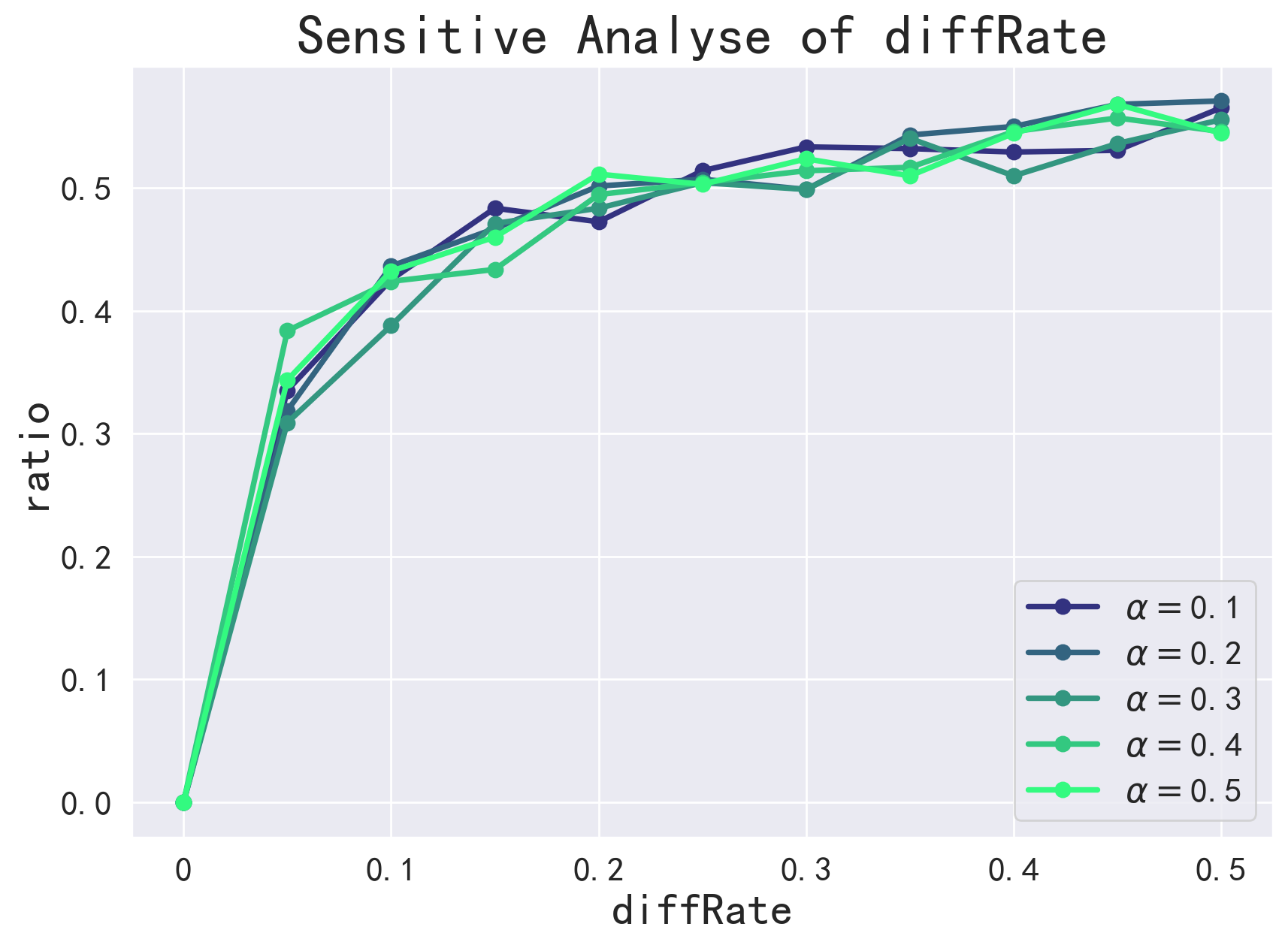}
		\caption{Sensitivity analysis for $\alpha$ and $diffRate$}
		\label{fig:sensitive analysis for propagation rate and acceptance rate}
	\end{minipage}
\end{figure} 

\autoref{fig:sensitive analysis for inter-layer coefficients and acceptance rate} and \autoref{fig:sensitive analysis for propagation rate and acceptance rate} exhibit a ratio of insusceptible attitudes under different inter-layer coefficients $\gamma_1$, $\gamma_2$ and acceptance rate $\beta$ when the propagation reaches stability. With the acceptance rate $\beta$ rising, the ratio of insusceptible states shows a noticeable upward trend. When the acceptance rates $\beta$ change in a low range, the simulation's final distribution variation will be significant. For instance, when the acceptance rate $\beta$ changes to a high-level range of around 0.4, the influence on the simulation system's final distribution will be slight. For inter-layer coupling effects between double-layer networks, on the other hand, the impact of inter-layer coefficients $\gamma_1$ and $\gamma_2$ can be ignored, meaning this parameter is robust for the entire system. Likewise, in the user association layer, the sensitivity of propagation rate $\alpha$ is low.
\begin{figure}[H]
	\centering
	\begin{minipage}[b]{0.58\textwidth}
		\includegraphics[width=\textwidth]{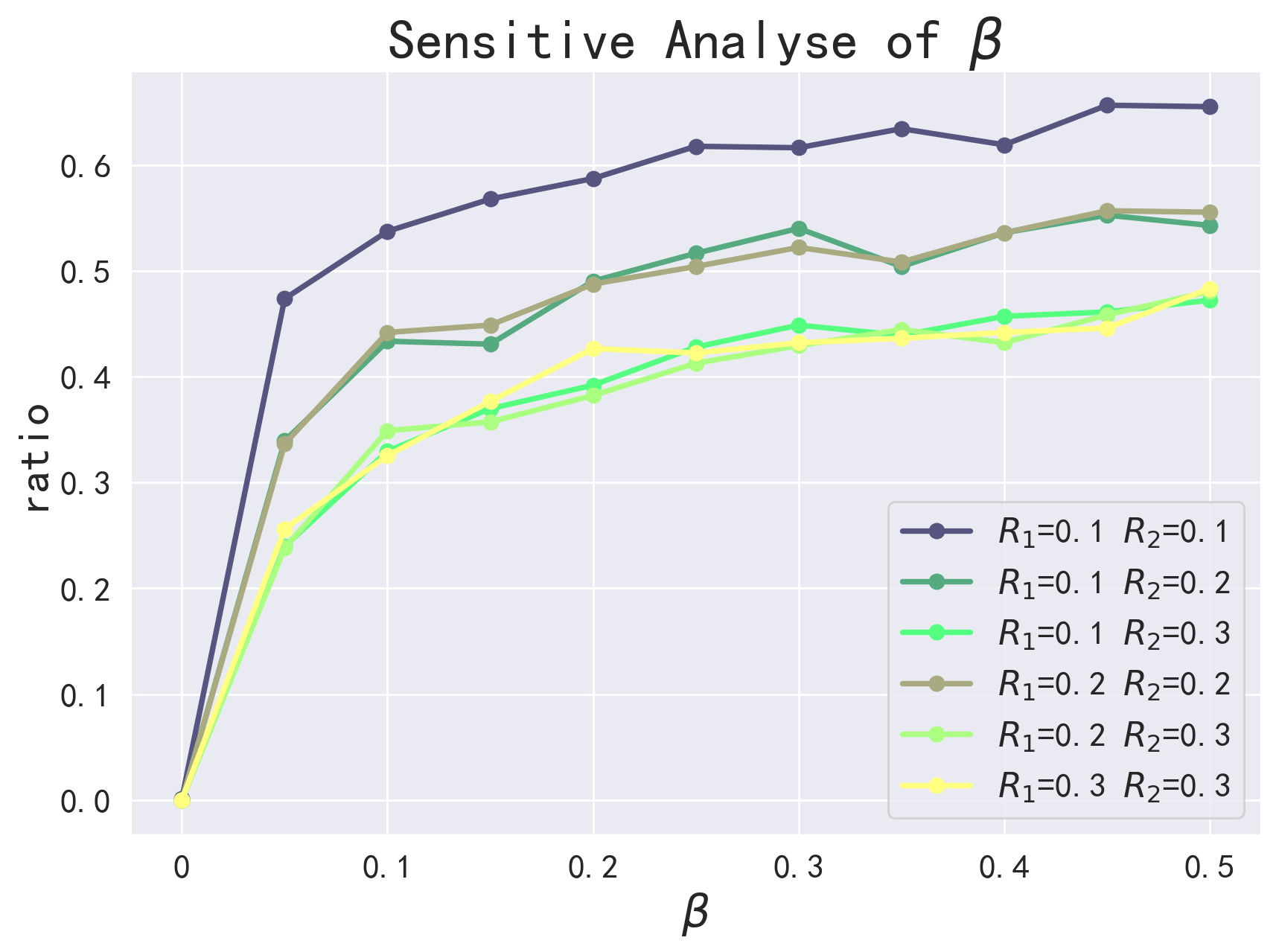}
		\caption{Sensitivity analysis for $R_1$, $R_2$ and $\beta$}
		\label{fig:sensitive analysis for acceptance rate, recover rate and transition rate}
	\end{minipage}
\end{figure} 

\begin{figure}[H]
	\centering
	\begin{minipage}[b]{0.58\textwidth}
		\includegraphics[width=\textwidth]{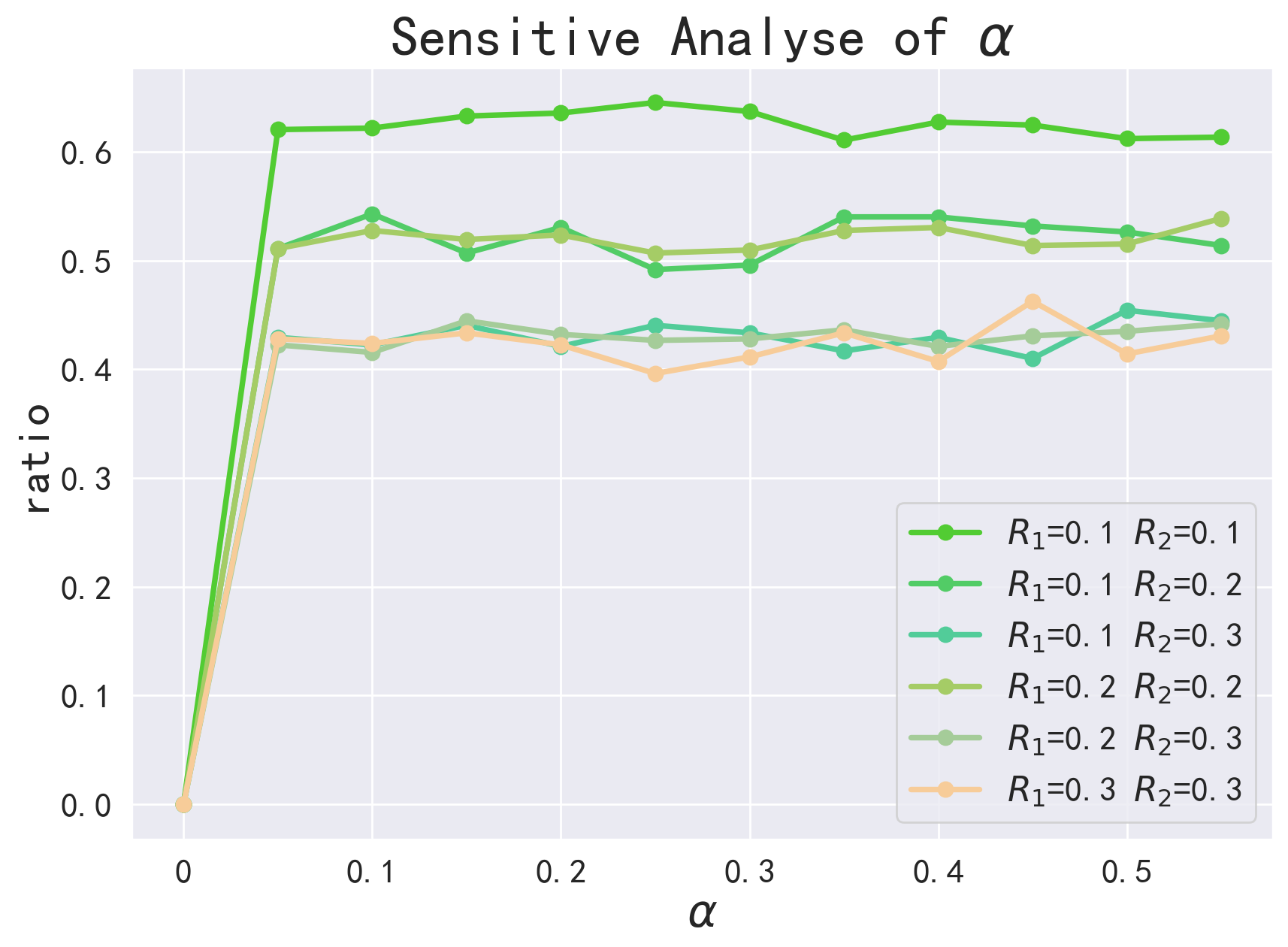}
		\caption{Sensitivity analysis for $R_1$, $R_2$ and $\alpha$}
		\label{fig:sensitive analysis for propagation rate, recover rate and transition rate}
	\end{minipage}
\end{figure} 
\autoref{fig:sensitive analysis for acceptance rate, recover rate and transition rate} and \autoref{fig:sensitive analysis for propagation rate, recover rate and transition rate} exhibit the sensitivity varying with propagation rate $\alpha$, acceptance rate $\beta$, recover rate $R_1$ and transition rate $R_2$. The transition rate $R_2$ in the similarity network has a direct impact on the simulation system: Namely, the nodes in the similarity networks transfer their states from susceptible to insusceptible. 

\subsection{Parameters Analysis}
As we can see from \autoref{fig:sensitive analysis  in community1} and \autoref{fig:sensitive analysis in community2}, when the attitudes of high influential nodes in the first community change, the users' attitudes distribution in the second community has little change while the attitudes in the first community change significantly, meaning that our novel community detection algorithm has fantastic sub-network segmentation effects. The sub-networks formed by each community exhibit relative close, wherein the influence of viewpoints among community is low and information exchange is relatively little. However, viewpoint exchange is mainly centralized within each sub-network, which also illustrates that the standpoints of individuals are easily influenced by users with high relevance with them.

For initialization parameters, intervention effects is sensitive to susceptible parameters $\theta$ and intervention ratio $\eta$, illustrating that the distribution of users' attitudes is related to the intensity of intervention measure implementation. Meanwhile, the level of acceptance plays an indispensable role in information cocoons emergence and extinction. 
This also confirms some conclusions about information cocoons drawn from scholars the improvement of literacy of the population in terms of compatibility with heterogeneous information and different opinions has profound significance in eliminating information cocoons \cite{santos2023break}.

For simulation system parameters, propagation rate $\alpha$ has little impact on our model, because such parameter has coupling effect with the level of users' acceptance to heterogeneous information, users' susceptible status and the number of neighbours holding heterogeneous attitudes. Thus, the level of propagation for each individual has little impact on the ultimate attitude distribution. Acceptance rate $\beta$ for individuals and recover rate $R_1$ have influence on sensitivity, meaning that states transition parameter has influence on users' sensitivity. To be precise, the tendency of users changing their states by adopting heterogeneous perspectives have significant influence on the dissemination of viewpoints and the formation of information cocoons.

\section{Conclusion}\label{section: Conclusion}
Although the recommendation system makes people's lives more convenient, its excessive use is widely reckoned to be the attribution of information cocoons and might contribute to group polarization. Therefore, it is necessary to regularize recommendation rules and design an automatic community detection monitor scheme. Consequently, it can precisely locate the key nodes that influence a small group of sub-networks at the initial stage of group polarization and adjust recommendation regulations on time, which could provide a more comprehensive understanding of the whole events for viewers, avoiding the occurrence of group polarization. In short, the problem brought by the algorithm can be solved by the algorithm itself through reasonable utilization. 

Aiming at quantity analysis for information cocoons, this paper proposes a novel multi-layer network community detection algorithm, which is effective for monitoring information cocoons. Simultaneously, this paper also proposes an intervention strategy in which algorithms can operate. The dissemination principle-based double-layer network Markov transition models are used to simulate intervention measures, verifying that our intervention strategy in this paper specifically affects de-homogenization for individuals within relatively closed sub-networks. Consequently, it illustrates that the intervention measure proposed in our paper can relieve the information cocoon phenomenon and meanwhile ensure members in another community are less affected. 

Some weaknesses exist in our models. Firstly, feature vectors are excavated only from word frequency without consideration of semantic perspectives. Due to the limitation of the social network dataset, we cannot obtain more users' attribution data. Thus, the second layer network is merely constructed by word weighted vectors. 
Secondly, apart from the three attitudes of positive, neutral, and negative, the node labels can be divided into different themes based on content and semantics.
Finally, the reasonable range of threshold values for the information cocoons monitoring scheme has not been proposed in experimental analysis. It is always based on experience when determining whether we need to exert intervention measures on sub-networks.

Further research would consider more user attribution, such as user attribution characteristics, users' history information, and semantic features. This could bring the similarity layer closer to recommendation regulations, thus making the segmentation of sub-networks more precise. Combining semantic information and user history attribution, we can narrow down the scope of this intervention and achieve precise determination indexes. 








\bibliographystyle{plain}
\bibliography{main.bib}

\appendix 
\section{Comparison Algorithm: TWIST}\label{TWIST}
The comparison algorithm tucker decomposition with integrated SVD transformation (TWIST) is based on the stochastic block model. 
\subsection{Single Layer Graph Generated Model}
Stochastic block model has been proposed since 1983 \cite{holland1983stochastic}. It has become the most common method in graph generation model, usually applied to community partition. Its essence is a probability generation model, taking the adjacent matrices as samples for estimating the probability of a link between two nodes. In general, a link is more likely to exist between two nodes that belong to the same community. In contrast, the nodes belong to different communities has less possibility to be associated. We can establish a probability graph representation matrix, thus deducing the community affiliation status.  

For graph $\mathcal{G}=\left(\mathcal{V},\mathcal{E} \right) $, the adjacent matrix is $A=(a_{ij})_{n\times n}$ and matrix $P=(p_{ij})_{n\times n}$ represents the probability of node $i$ and node $j$ belonging to the same community. One-hot vectors represent community labels. Assume we have K communities in total and the community representation vector is noted as $Z_i=\left(z_{i1},\cdots,z_{im},\cdots,z_{iK} \right)^T\in\mathbb{R}^K $, where $z_{im}$ is the $m$-th entities of vector $Z_i$ with a value of 0 or 1. Graph representation matrix $Z=\left(Z_1^T,\cdots,Z_p^T,\cdots,Z_n^T \right)^T\in\mathbb{R}^{n\times K}$ with community information is composed of graph representation vector for each node. Denote the probability $\pi_m$ is $i$-th node affiliating with the $m$-th community, namely $\mathbb{P}\left(z_{im}=1 \right)=\pi_m $ and $\sum_{m=1}^{K}\pi_m=1.$ Stochastic block model assume that the probability of existing an edge between two nodes obeys $a_{ij}\sim Bernoulli\left(1,Z_i^TPZ_j \right). $

\subsection{Multi-Layer Graph Generated Model}
In multilayer networks, huge differences exist between each layer, so the task for community detection is finding a suitable partition including features in each layer. Unlike the stochastic community setting in single layer networks, the multi-layer networks stochastic block model assumes each layer randomly belonging to a special partition that the stochastic model determines. Assume there is an $L$ layers network with $M$ different community partition. Denote the $l$-th layer belongs to partition $\psi_l$. The probability of $l$-th layer affiliates community $m$ is marked as $\pi_m$, then $\mathbb{P}\left(\psi_l=m \right)=\pi_m\;(l=1,2,\cdots,L)$ and $\sum\limits_{m=1}^{M}\pi_m=1.$

A single layer stochastic block model generated the $m$-th partition. The probability matrix $P_m=(p_{ij})_{K_m\times K_m}\in\mathbb{R}^{K_m\times K_m}$ with the value between 0 and 1 represents the probability of communities with association. Community affiliation indicator matrix $Z_m\in\mathbb{R}^{n\times K_m}$ with one-hot vectors constituting each row represents which community each node belongs to. Similarly, the existence of a link between node $i$ and node $j$ obeys Bernoulli distribution. For the $l$-th layer network $(1\le l\le L)$, the probability of an edge between node $i$ and node $j$ follows Bernoulli distribution: $a_{ij}^l\sim Bernoulli\left(Z_{\psi_l}(i,:)P_{\psi_l}Z_{\psi_l}(j,:)^T \right).$ Thus, the $l$-th layer of adjacent tensor $A^l$ is: $\mathbb{E}\left(A^l|\psi_l\right)=Z_{\psi_l}(i,:)P_{\psi_l}Z_{\psi_l}(j,:)^T. $

\subsection{Tensor Decomposition}
For each single layer network $\mathcal{G}_l$ in a $L$ layer network $\mathcal{G}=\left\lbrace\mathcal{G}_1,\cdots,\mathcal{G}_L \right\rbrace $, there exists $n\times K_l$ dimensional graph representation matrix. The global graph representation matrix $Z=\left(Z_1,Z_2,\cdots,Z_m \right)\in\left\lbrace0,1 \right\rbrace^{n\times\sum_{j=1}^{L}K_j}  $ is obtained by concatenating each layer of graph representation matrix.  

From Tensor Representation Theorem \cite{jing2021community}, $\mathbb{E}(A|L)=P\times_1C\times_2C\times_3R.$, where $R$ is layer partition matrix, each line of which is composed of a one-hot vector. $R=(\gamma_{\psi_1},\gamma_{\psi_2},\cdots,\gamma_{\psi_L})^T\in\left\lbrace0,1 \right\rbrace^{L\times M} $ means each layer belongs to which kind of partition.

\subsection{Adjacent Tensor Decomposition and Graph Representation Tensor Estimation}
We apply SVD decomposition to community representation matrix: $C=Z\Sigma V^T$, where $Z\in\mathbb{R}^{n\times r}$ and $V\in\mathbb{R}^{\sum_{j=1}^{M}K_j\times r}$. Matrix $\Sigma$ is composed of community representation matrix singular value in descending order:
\begin{align}
	\Sigma=\left[\begin{matrix}
		\sigma_1(\Sigma)&\cdots&0\\
		\vdots&\ddots&\vdots\\
		0&\cdots&\sigma_r(\Sigma)
	\end{matrix} \right] ,
\end{align}  
where $\sigma_1(\Sigma)\ge\sigma_2(\Sigma)\ge\cdots\ge\sigma_r(\Sigma) $. Then 
\begin{align*}
	\mathbb{E}\left(A|L\right)=P\times_1C\times_2C\times_3R=S\times_1Z\times_2Z\times_3R' ,
\end{align*}
where the tensor $S=P\times_1(\Sigma V^T)\times_2(\Sigma V^T)\times_3D^\frac{1}{2}\in\mathbb{R}^{r\times r\times M}$ and $D=R'^{-1}(RR)R'^{-1}$.

We can determine whether node $i$ and node $j$ belongs to the same community through the distance between their representation vector. If node $i$ and node $j$ are not in the same community, then $\left\|Z(i,:)-Z(j,:)\right\|\ge\frac{1}{\max(\sigma(\Sigma))}$.

\begin{figure}[H]
	\centering
	\begin{minipage}[b]{0.6\textwidth}
		\includegraphics[width=\textwidth]{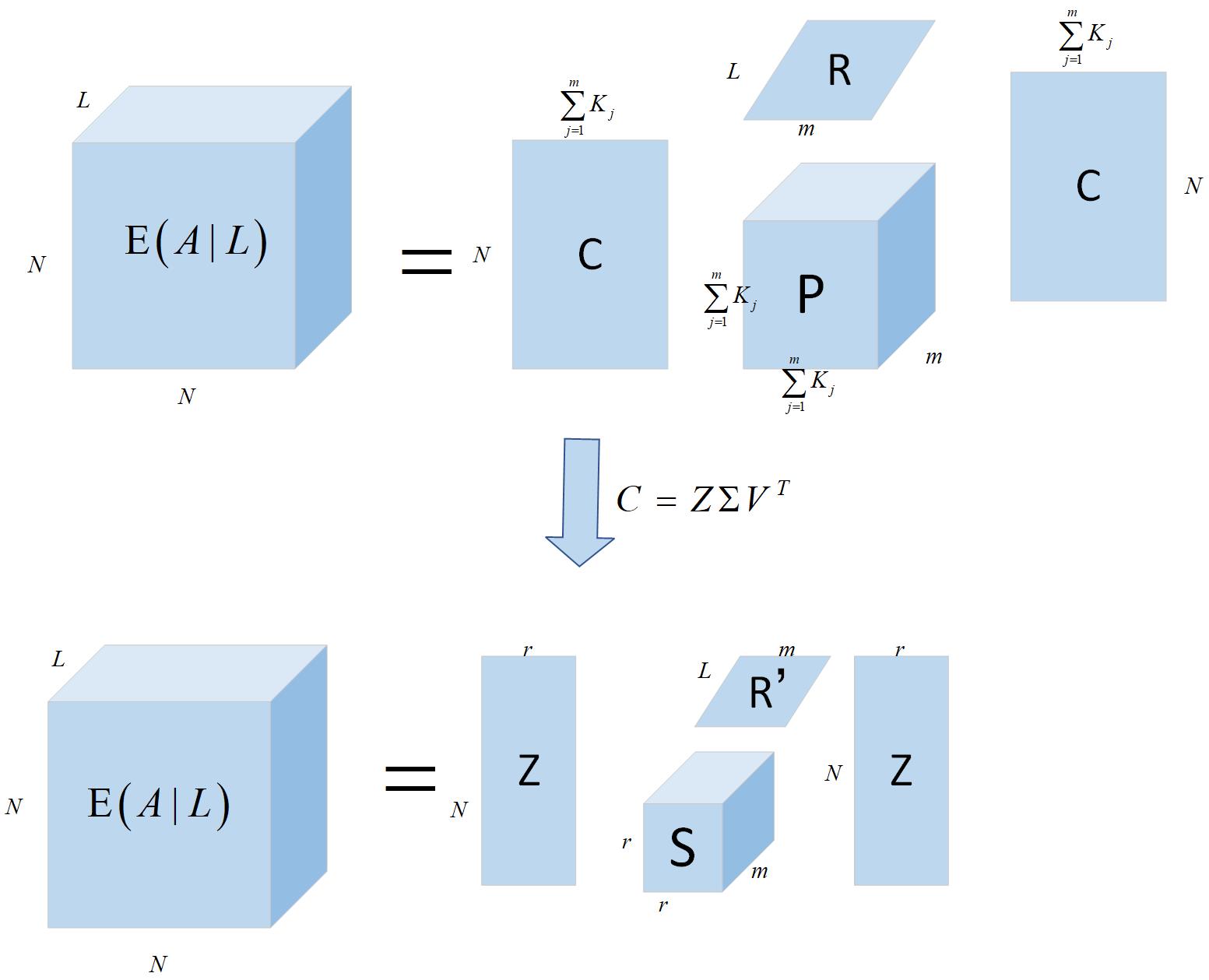}
		\caption{TWIST Algorithm}
		\label{fig:TWIST}
	\end{minipage}
\end{figure}

\section{Lemmas}
\begin{lemma}\label{[Eckart and Young Theorem]}
	(Eckart and Young Theorem \cite{eckart1936approximation}) For a real matrix $A\in\mathbb{R}^{m\times n}\;(n\ge m)$. Suppose the singular value decomposition of $A$ is $A=U\Sigma V^T$. The optimal solution of the following minimization problem \eqref{best r rank approximation} 
	\begin{align}
		\arg\min\limits_{\substack{{\hat{A}\in\mathbb{R}^{m\times n}}\\ {rank(\hat{A})\le r}}}
		\left\|A-\hat{A}\right\|_F
	\end{align}\label{best r rank approximation}
	is called the best r rank approximation of $A$, which is formulated as $\hat{A}_k=\sum\limits_{i=1}^{r}\sigma_i u_i v_i^T$. And 
	\begin{align*}
		\left\|A-\hat{A}_k\right\|_F=\sqrt{\sigma_{r+1}^2+\cdots+\sigma_{n}^2}.
	\end{align*}
	
\end{lemma}

\begin{lemma}\label{low rank approximation}
	[Low Rank Approximation] The modularity maximization problem \eqref{modularity maximization problem} in single layer network 
	\begin{align}
		\max\limits_{Tr(Z^TZ)=N}Tr\left(Z^TBZ \right)
	\end{align}\label{modularity maximization problem}
	is equivalent to $r$th-order low-rank reconstruction of modularity degree matrix $B$. Moreover, the $r$th-order low-rank reconstruction can be formulated as $\hat{B}_r=\Psi^T\Psi$, where $\Psi=H\Lambda$, $\Sigma_r=\Lambda^T\Lambda$ and $B=[H,P]\Sigma[H,P]^T.$
\end{lemma}

\begin{proof}[Proof of \autoref{[Eckart and Young Theorem]}]
	The singular value decomposition of $A$ is $A=U\Sigma V^T$.
	\begin{align*}
		&\left\|A\right\|_{F}=\left\|U\Sigma V^T\right\|_{F}=\sqrt{Tr(V\Sigma^TU^TU\Sigma V^T)}=\sqrt{Tr(V\Sigma^T\Sigma V^T)}\\
		=&\sqrt{Tr(\Sigma^T\Sigma V^TV)}=\sqrt{Tr(\Sigma^T\Sigma)}=\sqrt{\sigma_1^2+\cdots+\sigma_n^2}.
	\end{align*}
	Thus 
	\begin{align*}
		\left\|A-\hat{A}_r\right\|_{F}=\sqrt{\sigma_{r+1}^2+\cdots+\sigma_n^2}.
	\end{align*}
    The proof is then finished.
\end{proof}

\begin{proof}[Proof of \autoref{low rank approximation}]
	We use the eigenvalue decomposition: $B=\left[\begin{matrix}
		H&P
	\end{matrix}\right]\left[\begin{matrix}
		\Sigma_r&O\\
		O&\Sigma_{n-r}
	\end{matrix}\right]\left[\begin{matrix}
		H^T\\P^T
	\end{matrix}\right]$,
	where $\Sigma_r$ is diagonal matrix composed of the top $r$-th eigenvalue.
	Consequently, the optimal $r$-rank approximation of modularity matrix B is $\hat{B}_r=H\Sigma_r H^T$. 
	From \autoref{[Eckart and Young Theorem]}, we know that the $r$-th largest eigenvalue of modularity tensor $B$. Thus, part of block matrix $\Sigma_\tau$ is positive-definite. 
	Let $\Sigma_r=\Lambda\Lambda^T$, then $\hat{B}_r=H\Lambda\Lambda^TH^T=(H\Lambda)(H\Lambda)^T.$
    
    The proof is then finished.
\end{proof}

\section{Computation of Influential Factors and Analysis}\label{Appendix: Influential Factors}
In this section, we formulate the details of iteration process for influential factors computation. 
\begin{algorithm}[H] 
	\caption{Computation of Influential Factors} 
	\label{alg:Framwork} 
	\begin{algorithmic}[1] 
		\STATE Initialize the parameters by degree centrality.\\
		$DC=[DC(1),DC(2),\cdots,DC(N)]^T,$\\
		$EC^{(0)}=DC,\;\lambda=0.85.$
		\STATE Calculate the normalization adjacent matrix in each layer $\tilde{A}^{(l)}=D+A$, where $D=diag\left( \sum_{j}a_{1j},\cdots,\sum_{j}a_{Nj}\right).$\\$\tilde{A}=\tilde{A}^{(1)}\cup\tilde{A}^{(2)}\cup\cdots\cup\tilde{A}^{(L)}$\\
		$M=\tilde{D}^{-1}\tilde{A}$
		
		\STATE Iteration:\\
		While $\left|EC^{(\alpha+1)}-EC^{(\alpha)}\right|>\epsilon\quad\&\quad\alpha<epoch:$\\
		$\qquad EC^{(\alpha+1)}=\lambda\cdot M\cdot EC^{(\alpha)}+(1-\lambda)EC^{(\alpha)}$\\
		$\qquad\alpha=\alpha+1$
	\end{algorithmic}
\end{algorithm}

To verify this algorithm, we test the top 100 of high influential users and visualize its maximum associated users. The maximum associated users is the cardinal number of maximum neighbour vertex: $\left|N_{max}(i)\right|=\left|\cup_{l=1}^{L}N(l,i)\right|$. 
\begin{figure}[H]
	\centering
	\begin{minipage}[b]{0.58\textwidth}
		\includegraphics[width=\textwidth]{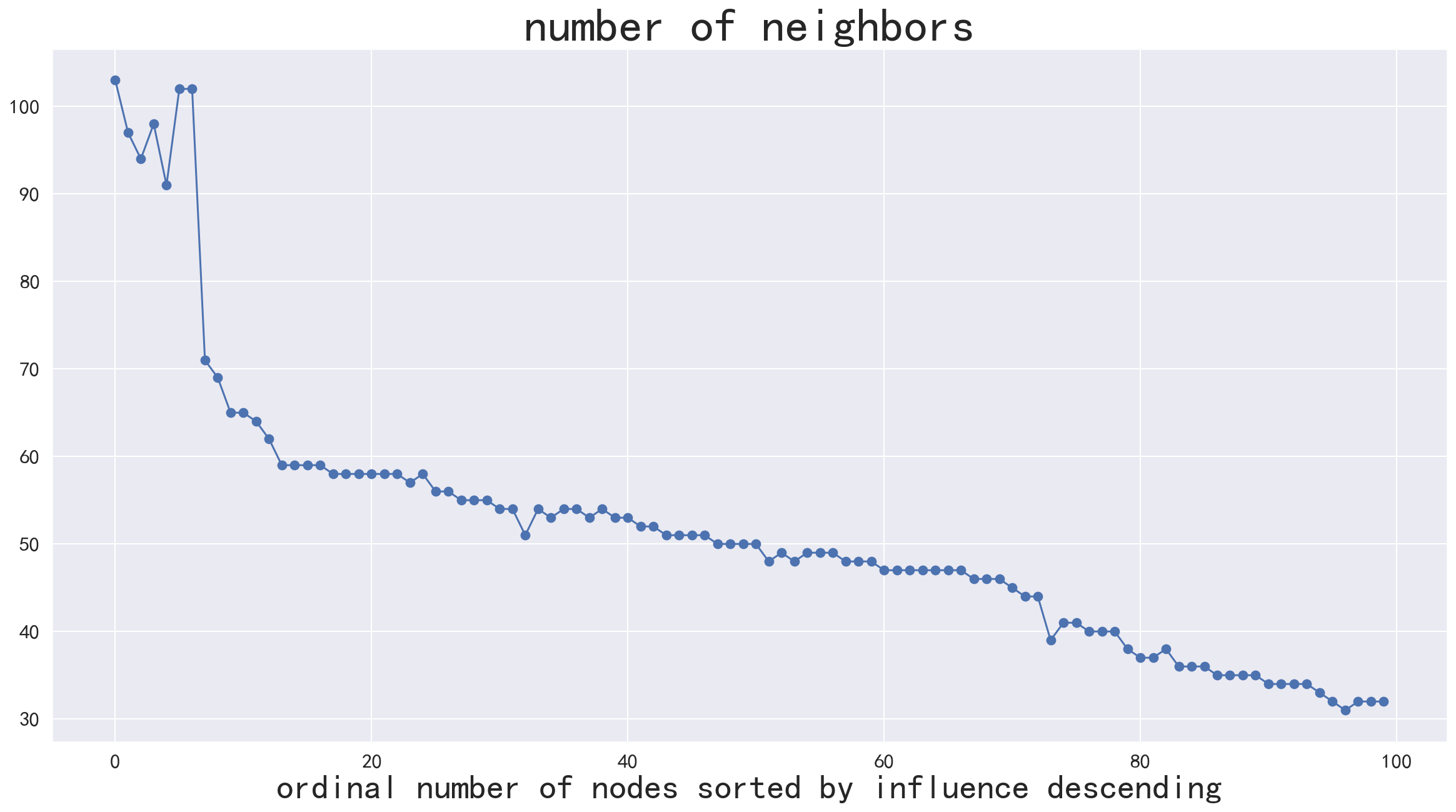}
		\caption{Neighbor count varying with influence rank.}
		\label{fig:the number of adjacent users varying with user' influential rank}
	\end{minipage}
\end{figure}
In \autoref{fig:the number of adjacent users varying with user' influential rank}, with the decrease of users' influential, the number of adjacent users appears corresponding downward trend except a slight fluctuation concentrating on that of high influential users.
\end{document}